\documentclass[a4paper,fleqn,usenatbib]{mnras}

\usepackage{savesym}
\usepackage{amsmath}
\savesymbol{iint}
\usepackage{txfonts}
\restoresymbol{TXF}{iint}

\usepackage[T1]{fontenc}
\usepackage{ae,aecompl}

\usepackage{graphicx}	
\usepackage{amsmath}	
\usepackage{amssymb}	
\usepackage{hyperref}

\title[Photospheric Monte Carlo Simulations]{Monte Carlo Simulations of the Photospheric Process}

\author[R. Santana et al.]{
Rodolfo Santana,$^{1}$\thanks{E-mail: santana@astro.as.utexas.edu (RS)}
Patrick Crumley,$^{1,2,3}$
Roberto A. Hern\'{a}ndez,$^{1,2}$
\& Pawan Kumar$^{1}$
\\
$^{1}$Department of Astronomy, University of Texas at Austin, Austin, TX, 78712, USA\\
$^{2}$Department of Physics, University of Texas at Austin, Austin, TX, 78712, USA\\
$^{3}$Astronomical Institute Anton Pannekoek, University of Amsterdam, P.O. Box 94249, 1090 GE Amsterdam, the Netherlands
}

\date{Accepted XXX. Received YYY; in original form ZZZ}

\pubyear{2015}

\begin{document}
\label{firstpage}
\pagerange{\pageref{firstpage}--\pageref{lastpage}}
\maketitle

\begin{abstract}
We present a Monte Carlo (MC) code we wrote to simulate the
photospheric process and to study the photospheric spectrum above the
peak energy. Our simulations were performed with a
photon to electron ratio $N_{\gamma}/N_{e} = 10^{5}$, as
determined by observations of the GRB prompt emission.
We searched an exhaustive parameter
space to determine if the photospheric process can match the observed
high-energy spectrum of the prompt emission. If we do not consider
electron re-heating, we determined that the best conditions to produce
the observed high-energy spectrum are low photon temperatures and high
optical depths. However, for these simulations, the spectrum peaks at an
energy below 300 keV by a factor $\sim 10$. For the cases we consider 
with higher photon temperatures and lower optical depths, we
demonstrate that additional energy in the electrons is required to produce a 
power-law spectrum above the peak-energy. By considering
electron re-heating near the photosphere, the spectrum for these
simulations have a peak-energy $\sim \mbox{300 keV}$
and a power-law spectrum extending to at least 10 MeV with a spectral
index consistent with the prompt emission observations.
We also performed simulations
for different values of $N_{\gamma}/N_{e}$ and determined that the
simulation results are very sensitive to $N_{\gamma}/N_{e}$.
Lastly, in addition to Comptonizing a Blackbody spectrum, we also simulate the
Comptonization of a $f_{\nu} \propto \nu^{-1/2}$ fast cooled
synchrotron spectrum. The spectrum for these simulations peaks at
$\sim 10^{4} \mbox{ keV}$, with a flat spectrum
$f_{\nu} \propto \nu^{0}$ below the peak energy.
\end{abstract}




\section{Introduction}

One of the major open questions in the GRB field is, what is the
radiation mechanism that
produces the observed gamma-ray spectrum, i.e. the Band
function \citep{band_et_al_1993}? The Band function is a non-thermal
broken power-law spectrum with the peak-energy typically observed at
$\sim \mbox{300 keV}$ \citep{kaneko_et_al_2006}. Below (above) the peak-energy, the observed
spectrum is a single power-law with a typical spectrum
$f_{\nu} \propto \nu^{-1.2}$ ($f_{\nu} \propto \nu^{0}$) in the energy
range $\sim$ 300 keV-10 MeV ($\sim$ 10 keV-300 keV) \citep{preece_et_al_2000}.
In order to explain the prompt emission, a radiation mechanism must be
able to explain all these in the observed spectrum.

The three main
mechanisms that have been studied to try and reproduce the Band spectrum are synchrotron, synchrotron self-Compton (SSC), and the photospheric
process \citep[See][for reviews on the GRB radiation mechanism]{piran_2004,zhang_2014,kumar_and_zhang_2014}. In
this work, we focus on the photospheric process
\citep[][]{goodman_1986,paczynski_1990,thompson_et_al_1994,ghisellini_and_celotti_1999,meszaros_and_rees_2000,
meszaros_et_al_2002, rees_and_meszaros_2005,thompson_et_al_2007},
which involves photons undergoing multiple
scatterings with hot electrons below the photosphere
(Comptonization). Although not a necessary condition for the
Comptonization of photons, studies on the photospheric process
typically consider the photons to initially have a Blackbody (BB)
spectrum, with the peak of the BB spectrum taken to match the observed
peak energy of the prompt emission. With an initial seed BB spectrum,
the goal of the photospheric
process is to broaden the BB spectrum so that it matches the observed Band
function. In this work, we focus on models of the photospheric process where the
broadening is due to hot
electrons scattering photons to higher energies multiple times,
where a baryon dominated jet 
\citep[][]{peer_et_al_2006,lazzati_and_begelman_2010,
toma_et_al_2011,lazzati_et_al_2013,chhotray_and_lazzati_2015},
a Poynting dominated jet
\citep[][]{giannios_2008,begue_and_peer_2015},
and a hydrid jet system with both baryons and magnetic fields
\citep{gao_and_zhang_2015} have all been considered.
Models of the photospheric process where the broadening is due to
geometrical effects have also been investigated
\citep[][]{peer_2008,peer_and_ryde_2011,mizuta_et_al_2011,ruffini_et_al_2013,begue_et_al_2013,lundman_et_al_2013,
ito_et_al_2014,begue_et_al_2014}. See \citealt{vereshchagin_2014} and
\citealt{peer_et_al_2015} for reviews on the photospheric process.

The basic picture of the photospheric process is as follows.
The photons are assumed to be produced below the photosphere. At an
optical depth $\tau \sim \mbox{ few}$ - 100, a dissipation event is assumed
to occur, which accelerates the electrons to mildly relativistic or
relativistic speeds\footnote{The dissipation events discussed in the literature for the
  photospheric process are magnetic
  reconnection \citep[][]{thompson_et_al_1994,giannios_and_spruit_2005,giannios_2006,giannios_2012}
  and internal shocks
  \citep[][]{daigne_and_mochkovitch_2002,lazzati_and_begelman_2010,toma_et_al_2011,lazzati_et_al_2013}.}.
In the photospheric process, the average energy of
the photons is taken to be much smaller than the average energy of the
electrons. Thus, while the outflow is still optically thick, the
photons and electrons undergo multiple scatterings,
and a photon gains energy from the electrons until its energy reaches the average
electron energy or until it escapes the photosphere.
The Comptonization of BB photons by hot electrons is
predicted to produce a power-law spectrum
above the BB peak because only a fraction
$f$ of the photons get scattered once by a hot electron to higher
energies, only a fraction $f ^{2}$ of the photons get scattered
twice by a hot electron to higher energies, and so on
\citep[][]{lazzati_and_begelman_2010,ghisellini_2013}.
Once the outflow reaches the photospheric radius, the medium becomes
optically thin and the photons escape the outflow. The resulting
observed spectrum is, a peak determined by the BB temperature of
photons, and it has a power-law above the peak energy\footnote{One of the major difficulties of the
photospheric process is reproducing the typically observed low-energy spectrum
$f_{\nu} \propto \nu^{0}$. In this work, we ignore the low-energy spectrum issue and focus on
the high-energy spectrum. For detailed discussions on the low-energy
index of the photospheric process, see
\citealt[][]{vurm_et_al_2013,lundman_et_al_2013,deng_and_zhang_2014}.}.

An important quantity needed to simulate the photospheric process is
the ratio of photons to electrons, which we now estimate for the
prompt emission.
The kinetic energy of the GRB jet is
$E_{KE} = N_{p} m_{p} c^{2} \Gamma$, where $N_{p}$ is the number of
protons. We consider the kinetic energy of the jet to be
carried primarily by protons. Then, taking most of the photons during the prompt
emission to have an energy near the peak energy of the spectrum
($E_{\mathrm{pk}}$), the energy radiated in gamma-rays is $E_{\gamma} = N_{\gamma} E_{\mathrm{pk}}$.
With these two expressions, we can calculate the ratio
of $E_{\gamma}$ to $E_{KE}$:
\begin{equation}
\frac{E_{\gamma}}{E_{KE}} = \frac{N_{\gamma}
  E_{\mathrm{pk}}}{N_{p} m_{p} c^{2} \Gamma} .
\end{equation}
Taking $N_{p} = N_{e}$ (if few electron-positron pairs are created in
the GRB jet, then $N_{p} \approx N_{e}$ due to charge neutrality) and
defining the efficiency in the conversion of kinetic energy of the
jet to prompt radiation as $\eta = E_{\gamma}/(E_{\gamma} + E_{KE})$,
we can solve for the photon to electron ratio
$N_{\gamma}/N_{e}$ \citep{chhotray_and_lazzati_2015}
\begin{equation}
\frac{N_{\gamma}}{N_{e}} = 10^{6} \left( \frac{\eta}{1-\eta} \right) \left( \frac{\Gamma}{300} \right)
\left( \frac{E_{\mathrm{pk}}}{\mbox{ 300 keV}} \right)^{-1} .
\label{photon_electron_ratio_eq}
\end{equation}
In this expression, we have normalized $\Gamma$ and $E_{\mathrm{pk}}$
to typical values. Taking an efficiency $\eta \sim 10\%$, the photon to electron ratio is
$N_{\gamma}/N_{e} \sim 10^{5}$. Thus, a ratio of photons to electrons $\sim 10^{5}$ is
required to simulate the photospheric process
\citep{lazzati_and_begelman_2010,chhotray_and_lazzati_2015}.

Previous MC photospheric works have demonstrated that there is a power-law above the peak
of the spectrum for $N_{\gamma}/N_{e} \sim 10^{1} - 10^{4}$, where
$N_{\gamma}$ ($N_{e}$) is the number of photons (electrons) considered
\citep{lazzati_and_begelman_2010,chhotray_and_lazzati_2015}. However,
whether the power-law above the peak of the spectrum is a robust
feature of the photospheric process is still in question since
realistic simulations with $N_{\gamma}/N_{e} = 10^{5} $ have not
been performed. We developed a new MC photospheric code capable
of performing simulations for realistic GRB $N_{\gamma}/N_{e}$
ratios. In this work, we present results for MC photospheric
simulations with $N_{\gamma}/N_{e} = 10^{5} $ for the first time.
We also perform an exhaustive parameter space search
to determine if the photospheric process
can produce the observed high-energy spectral index of the Band
function.
In addition, we include adiabatic cooling of photons and
electrons, which was neglected by previous MC photospheric codes
\citep{lazzati_and_begelman_2010,chhotray_and_lazzati_2015}.

Another possible source for the seed photons is the synchrotron
process. Therefore, in addition to considering the Comptonization of a BB
spectrum of photons, we also consider the
Comptonization of a seed photon spectrum $f_{\nu} \propto \nu^{-1/2}$,
the expected synchrotron spectrum when electrons are in the fast
cooling regime \citep{sari_et_al_1996,ghisellini_et_al_2000}.
We use our MC photospheric code to study how
Comptonization modifies this seed spectrum.

This work is organized as follows. In Section \ref{algorithm_sect},
we describe the algorithm of our MC photospheric code (the expressions
for the implementation can be found in the Appendices). In Section
\ref{paramameter_search_sec}, we discuss the parameters we explore for our
MC photospheric simulations with a seed BB spectrum. The simulation
results for the Comptonization of a seed BB spectrum are discussed in Section
\ref{results_and_interpretation_sect} and the interpretation of these
results is discussed in Section \ref{discussion_of_results}.
In Section \ref{sync_comp_sect}, we discuss the parameters we explore
for our simulations on the Comptonization of a seed
$f_{\nu} \propto \nu^{-1/2}$ spectrum, the results for
these simulations, and the interpretation of these results. Lastly, in Section
\ref{conclusions_sect}, we discuss our conclusions.
%
%
%

\section{Description of Monte Carlo Photospheric Code}
\label{algorithm_sect}

In this section, we give an overview of our MC
photospheric code algorithm. The details for implementation can be
found in the Appendices. Our code was written in the C$++$11 programing
language and we used the GCC version 4.9.2 compiler. Under 9
GB of RAM are needed
for a simulation with $10^{8}$ photons and a simulation initialized at
$\tau_{\mathrm{initial}} = 2$ takes under 2 hours in a regular desktop Linux
machine (see definition of $\tau_{\mathrm{initial}}$ below). Lastly, we note that
our code is not parallelized; each scattering event between a photon
and an electron is performed one by one.
\subsection{Input Parameters for Simulations with Seed BB Photons}
\label{input_param_sect}

The input parameters for our MC photospheric simulations with a seed
BB spectrum of photons are described below. In Section
\ref{sync_comp_sect}, we will discuss the input parameters for the
simulations with a $f_{\nu} \propto \nu^{-1/2}$ seed spectrum.
In the discussion below and throughout this work, unprimed (primed) quantities
refer to quantities in the observer (jet-comoving) frame.
\begin{itemize}
\item $\Gamma$ --- The bulk Lorentz factor of the
  outflow. We consider a typical $\Gamma = 300$ for GRBs
\citep{molinari_et_al_2007,xue_et_al_2009,liang_et_al_2010}.
\item $L$ --- The isotropic equivalent kinetic luminosity
  of the outflow. We consider
  $L = 10^{52} \mbox{ ergs}/\mbox{sec}$
  \citep{liang_et_al_2007,wanderman_and_piran_2010}.
\item $N_{e}$ --- The number of electrons in a simulation.
We consider $N_{e} = 10^{3}$, the same number of electrons
as previous MC photospheric simulations
\citep{lazzati_and_begelman_2010,chhotray_and_lazzati_2015}.
In Section \ref{mc_code_tests}, we explicitly show that $10^{3}$
electrons are enough to get an accurate representation for a electron
distribution.
\item Electron Distribution --- We consider three different
  distributions for the electrons:
1. mono-energetic electrons (all electrons initialized to the same
electron Lorentz factor $\gamma_{e} ^{\prime}$),
with the initial $\gamma_{e} ^{\prime}$ of the electrons as the input
parameter. 2. Maxwell-Boltzmann (MB)
distribution of electrons with the electron temperature
$T ^{\prime}_{e}$ as an input parameter. 3. Power-Law
distribution of electrons $d N_{e}/d \gamma_{e} ^{\prime} \propto
(\gamma_{e} ^{\prime})
^{-p}$ ranging from $\gamma_{e, 1} ^{\prime}$ to
$\gamma_{e, 2} ^{\prime}$, with the electron Lorentz factors
$\gamma_{e, 1} ^{\prime}$, $\gamma_{e, 2} ^{\prime}$,
and the electron index $p$ as input parameters.
\item $\tau_{\mathrm{initial}}$ --- The optical depth corresponding to
  the distance from the central engine where the photons are
  initialized (see Equation \ref{R_obs_frame_eq}). We consider
 $\tau_{\mathrm{initial}} = $ 2, 8, 16.
\item $N_{\gamma}$ --- The number of photons in a simulation.
Since we typically consider $N_{e} = 10^{3}$, to reach
 $N_{\gamma}/N_{e} = 10^{5}$, we
consider $N_{\gamma} = 10^{8}$ for our simulations.
\item $T_{\gamma} ^{'}$ --- The photons are initialized to have a
  Blackbody (BB) distribution with temperature $T_{\gamma} ^{'} $.
\item $N_{\mathrm{collect}}$ --- The number of photons collected for
  the output spectrum. We consider $N_{\mathrm{collect}} =
  N_{\gamma}/3$ for our simulations as in
  \citealt{lazzati_and_begelman_2010} since considering $N_{\mathrm{collect}} =
  N_{\gamma}/3$ allows for enough scatterings to occur so that the
  electrons can cool by IC scatterings. By plotting the first
  $N_{\gamma}/3$ photons that escape the photosphere for an output
  spectrum, we are plotting a time-averaged spectrum.
\end{itemize}

\subsection{Initializing Electrons and Photons}

The first step of our MC photospheric simulations is to initialize the
electrons and photons. The only property we track
in the observer frame is the position of the photons to determine if
they have escaped the photosphere. All the other properties and
calculations are done in the jet-comoving frame.

\subsubsection{Initialization of Direction and Energy of Electrons}

The directions
of the momentum of the $N_{e}$ electrons are drawn randomly in the
jet-comoving frame (see Appendix \ref{appendix_elec_dir} for algorithm).
The $\gamma_{e} ^{\prime}$ for each of the $N_{e}$ electrons is drawn
from the distribution specified in the input parameters
(see Appendix \ref{appendix_elec_gam_e_MB_PL} for algorithm). We
assume that the electrons are distributed uniformly in the jet and do
not track their position.

\subsubsection{Initialization of Direction, Energy, and Position of Photons}

The directions of the momentum of the $N_{\gamma}$ photons are drawn randomly in the
jet-comoving frame (see Appendix \ref{appendix_phot_dir} for algorithm).
The energy of each of the $N_{\gamma}$ photons in the jet-comoving
frame ($E_{\gamma} ^{\prime}$) is drawn from either a BB
distribution with temperature $T_{\gamma} ^{'}$ or a power-law
distribution (depending on the system being investigated)
(see Appendix \ref{appendix_phot_ener} for algorithm).

The origin of the coordinate system we use to track the position of
the photons in the observer frame is the central engine.
The $N_{\gamma}$ photons are initially placed randomly and uniformly
within an angle  $ \le 1/\Gamma $ (jet opening angle) in the
direction of the observer at a distance
\begin{equation}
R_{\mathrm{initial}} = \frac{L \sigma_{T}}{8 \pi m_{p} c^{3} \beta \Gamma^{3} \tau_{\mathrm{initial}}} ,
\label{R_obs_frame_eq}
\end{equation}
where
$\sigma_{T}$ is the Thomson cross section
and $\beta = \sqrt{1 - \Gamma^{-2}}$ is the speed of the outflow
divided by the speed of light.

We then draw the distance $s^{\prime}$ each photon travels
in the jet-comoving frame before running into an electron
randomly from the probability density
$p(s^{\prime}) \propto \mathrm{exp}(-s^{\prime}/\ell_{\mathrm{mfp}} ^{'})$,
where $\ell_{\mathrm{mfp}} ^{\prime}$ is the mean free path. Inverting this probability density, $s^{'}$ is
sampled with the formula
$s^{\prime} = - \ell_{\mathrm{mfp}} ^{\prime} \mbox{ln}(\xi) $,
where $\xi$ is a uniformly distributed random number between 0 and 1.
The mean free path and the electron density ($n_{e} ^{\prime}$)
are given by
\begin{eqnarray}
\ell_{\mathrm{mfp}} ^{\prime} (R) &=& \frac{1}{ n_{e} ^{\prime} (R) \sigma_{T}}
\label{mean_free_path_comoving} \\
n_{e} ^{\prime} (R) &=& \frac{L}{4 \pi R^{2} m_{p} c^{3} \Gamma^{2} } ,
\label{electron_density_comoving}
\end{eqnarray}
where $R$ is the distance of the photon from the central engine in the observer frame. The
distance each photon travels in the jet comoving frame is then Lorentz
transformed to the observer frame to determine the new location of the
photon in the observer frame (see Appendix \ref{appendix_phot_propagation} for algorithm).

\subsection{Adiabatic Cooling of Photons and Electrons}

As the jet expands outward, the energy of the photons and
electrons decreases due to adiabatic cooling. Adiabatic cooling
depends on the radial distance traveled by the jet, with the energy
of the photons decreasing by a factor $r^{-2/3}$ and the kinetic
energy of the electrons
decreasing by a factor $r^{-2/3}$ or $r^{-4/3}$, depending on
whether a electron is relativistic or sub-relativistic. This scaling
is valid as long as the radial width of the jet does not change with
distance, which is satisfied for highly relativistic jets below the
photosphere. Thus, in between scattering
events, the expressions we use to update the energy of a photon and
the $\gamma_{e} ^{\prime}$ of a electron due to adiabatic cooling are
\begin{eqnarray}
\frac{E_{\gamma, f} ^{\prime}}{E_{\gamma, i} ^{\prime}} &=&
\left( \frac{R_{\mathrm{initial}} + (t_{\gamma} + \Delta t_{\gamma}) \beta_{j} c}
{R_{\mathrm{initial}} + t_{\gamma}  \beta_{j} c } \right)^{-2/3}
\label{adiab_cool_photons} \\
\frac{ \gamma_{e, f} ^{\prime} -1 }{ \gamma_{e, i} ^{\prime} -1 } &=&
\left( \frac{R_{\mathrm{initial}} + (t_{\gamma} + \Delta t_{\gamma}) \beta_{j} c}
{R_{\mathrm{initial}} + t_{e}  \beta_{j} c }
\right)^{2 - 2(4 \gamma_{e,i} ^{\prime} + 1)/(3 \gamma_{e, i} ^{\prime}  )   }
\label{adiab_cool_electrons} .
\end{eqnarray}
In these equations, $R_{\mathrm{initial}}$ corresponds to the
distance where the photons are initialized
(Equation \ref{R_obs_frame_eq}) and the subscript
$i$ $(f)$ corresponds to a photon and electron
property before (after) the photon travels a distance $s^{\prime}$
(in the jet comoving frame).  $t_{\gamma}$ ($t_{e}$)
represents the total time elapsed for a
photon (electron) in between scattering events (in the observer
frame). Thus, in the time $t_{\gamma}$ ($t_{e}$), the jet has traveled
a radial distance $t_{\gamma}  \beta_{j} c$ ($t_{e}  \beta_{j} c$)
and the radial position of the photon (electron) before the
photon travels a distance $s^{\prime}$ is
$R_{\mathrm{initial}} + t_{\gamma}  \beta_{j} c$
($R_{\mathrm{initial}} + t_{e}  \beta_{j} c$).
$\Delta t_{\gamma}$ represents the time it
takes a photon to travel a distance $s^{\prime}$ (in the observer
frame, see Appendix \ref{appendix_phot_propagation}); thus,
the final radial position of the photon and electron are
$R_{\mathrm{initial}} + (t_{\gamma} + \Delta t_{\gamma}) \beta_{j} c$. Lastly, we note that
the $(4 \gamma_{e,i} ^{\prime} + 1)/(3 \gamma_{e, i} ^{\prime} )$ term in the exponent of
Equation \ref{adiab_cool_electrons} is used to take into account that
the electron adiabatic index transitions from $4/3$ to $5/3$ as the
electron cools due to IC scatterings and adiabatic expansion.

\subsection{Main MC Photospheric Program}

The first step in our main program is to check for photons that were
able to escape the photosphere, without interacting with an electron,
with the first $s^{\prime}$ drawn. $R_{\mathrm{photosphere}}$ 
is defined as the radius where $\tau=1$ in Equation \ref{R_obs_frame_eq}. 
If a photon was able to escape the photosphere,
we Doppler boost its energy to the
observer frame with the Doppler factor,
$\mathcal{D} E_{\gamma} ^{\prime}$, and store this
energy. Otherwise, we place this photon in a priority queue data
structure. Each element in the priority queue is a tuple with entries
($t_{\gamma, k}$, $k$), where $k$ refers to the index of a photon in
an array and $t_{\gamma, k}$ refers to the total elapsed time in
between scatterings of this particular
photon (in the observed frame). With the photons in a priority queue,
the photon with the smallest $t_{\gamma, k}$ gets scattered first (is
given priority) and then the photon with the next smallest
$t_{\gamma, k}$ gets scattered next and so on \citep{lazzati_and_begelman_2010}.
Having the array index of the photon
($k$) allows us to access the properties of
this particular photon (energy, direction, and position).

In the next part of the program, we take into account adiabatic
cooling and determine whether a photon-electron scattering event will
occur. We first draw one of the $N_{e}$ electrons randomly. Then, we
use the $s^{\prime}$ of the first photon in the priority
queue (photon with smallest $t_{\gamma, k}$) to propagate this photon
forward (see Appendix \ref{appendix_phot_propagation} for algorithm) . After using
Equations \ref{adiab_cool_photons} - \ref{adiab_cool_electrons}
to take into account adiabatic cooling,
we calculate the dimensionless photon energy of this particular photon
in the rest frame of this particular electron (defined as $x_{i} ^{\prime}$)
and the cross-section for this interaction
(defined as $\sigma(x_{i} ^{\prime})$; see Appendix \ref{appen_elec_phot_scatt} for algorithm).
The probability that the scattering event will occur is $\sigma(x_{i} ^{\prime})/\sigma_{T}$ ,
where $\sigma(x_{i} ^{\prime}) \le \sigma_{T}$. To determine if the
scattering event occurs, we draw a uniformly
distributed random number between 0 and 1, defined as $\xi_{s}$.
If $\xi_{s} \le \sigma(x_{i} ^{\prime})/\sigma_{T}$, the scattering event
occurs. If the scattering event occurs, we update the energy and
direction of the photon
(see Appendix \ref{appen_elec_phot_scatt} for algorithm)
and the energy and direction of the electron
(see Appendix
\ref{appen_elec_ener_dir_update} for details)
after the scattering event.
Then (regardless of
whether the scattering event occurred or not), we draw a
new $s^{\prime}$ at the current location $R$ of the photon with the expression
$s^{\prime} (R) = - \ell_{\mathrm{mfp}} ^{\prime} (R) \mbox{ln}(\xi) $.
The distance the photon travels in the jet-comoving frame is then
Lorentz transformed to the observer frame to check if the photon has
escaped the photosphere. If
$R \ge R_{\mathrm{photosphere}}$ is satisfied, we store the energy of
this photon in the observer frame ($\mathcal{D} E_{\gamma} ^{\prime}$).
Otherwise, we place this photon back in the priority queue with the new
$t_{\gamma, k}$. We repeat the process described in this paragraph
until $N_{\mathrm{collect}}$ photons have escaped the photosphere.

\subsection{MC Photospheric Code Tests}
\label{mc_code_tests}

\begin{figure*}
\begin{center}$
\begin{array}{c}
\begin{array}{cc}
\hspace{-10mm}
\includegraphics[scale=0.45]{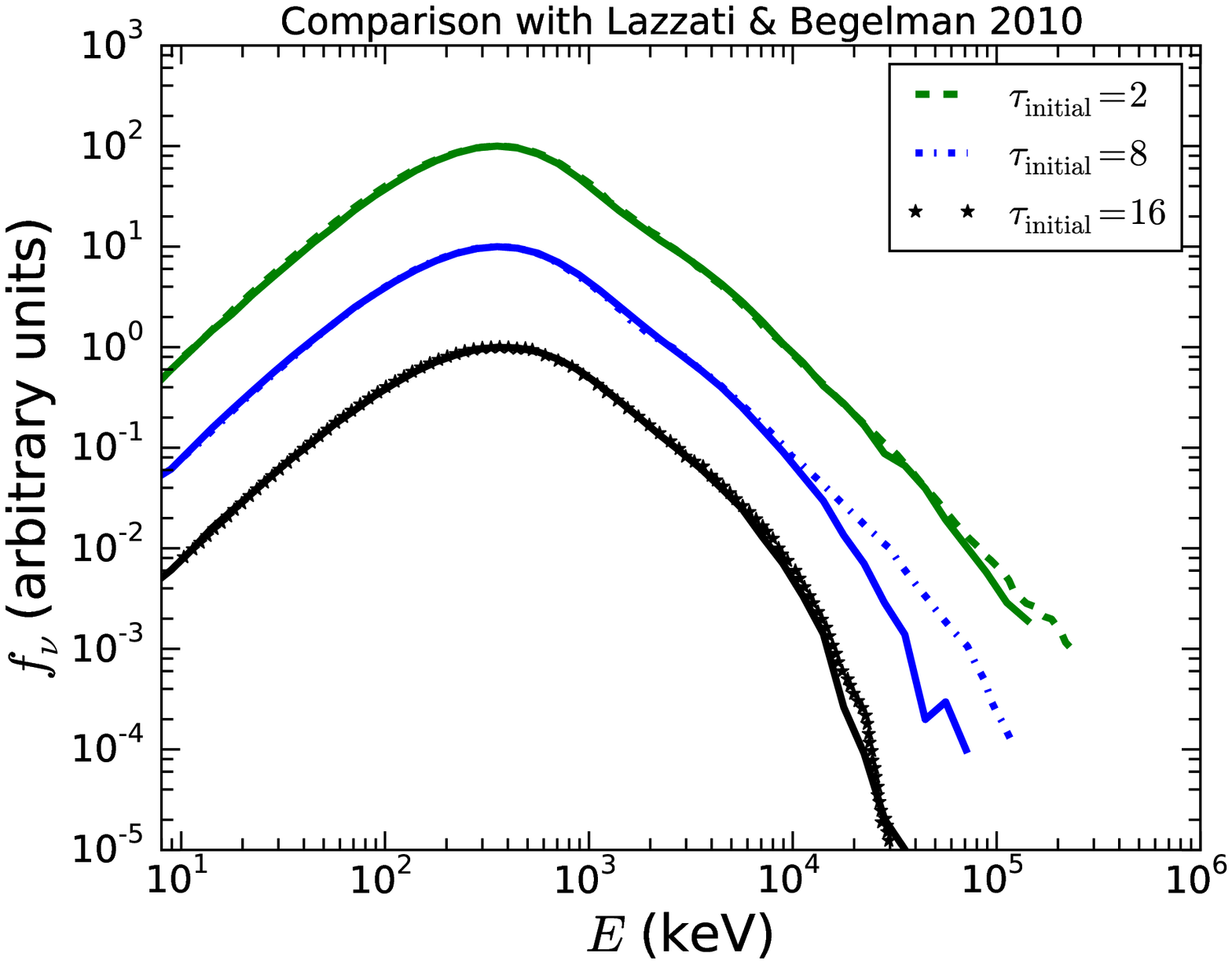} &
\hspace{0mm}
\includegraphics[scale=0.45]{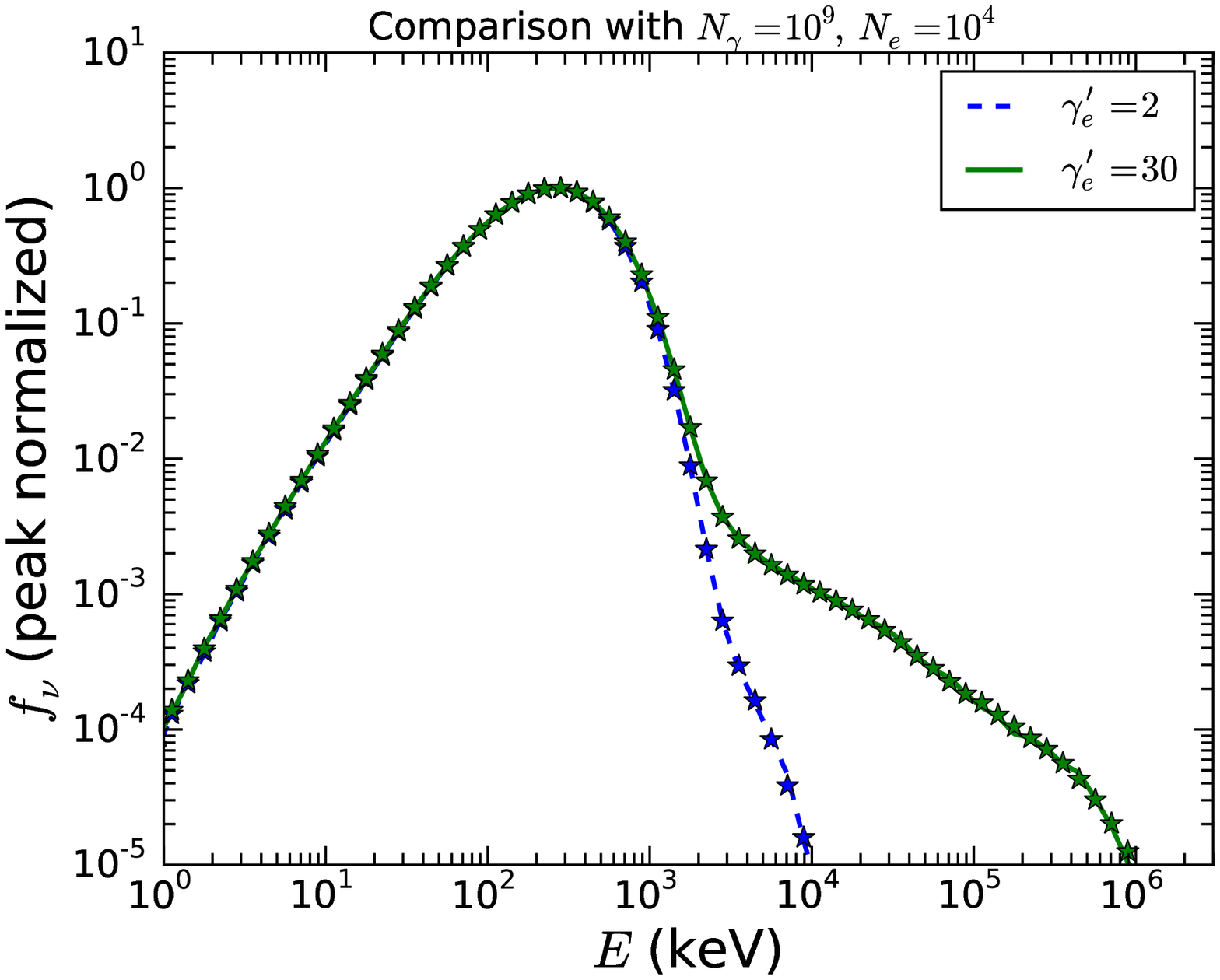}
\end{array}
\end{array}$
\end{center}
\caption{\textit{Left Panel:} Comparison of our MC photospheric simulation
results (solid lines) to those from Figure 4 of
\citet{lazzati_and_begelman_2010} (dotted lines, dash dotted lines,
and stars). These simulations are for the
Comptonization of seed BB photons with
$k_{B} T_{\gamma} ^{\prime} = \mbox{ 90 eV}$, $\Gamma=1000$ with
mono-energetic electrons with initial $\gamma_{e} ^{\prime} = 2$,
$\tau_{\mathrm{initial}} =$ 2, 8, 16, and no adiabatic cooling. \textit{Right Panel:}
Comparison of $N_{\gamma} = 10^{9}$, $N_{e} = 10^{4}$ (stars)
simulation results to $N_{\gamma} = 10^{8}$, $N_{e} = 10^{3}$ (dotted
and solid lines) simulation
results. The simulations are for the Comptonization of seed BB
photons with $k_{B} T_{\gamma} ^{\prime}=$ 300 eV,
$\Gamma=300$ with
mono-energetic electrons with initial $\gamma_{e} ^{\prime} =$ 2,
30, $\tau_{\mathrm{initial}}=2$, and adiabatic cooling. \label{fig1}}
\end{figure*}

To test our code, we compared our simulation
results to the MC photospheric code results from
\citet{lazzati_and_begelman_2010}. In the left panel of Figure \ref{fig1},
we compare our simulation results to the three simulations in Figure 4
of \citet{lazzati_and_begelman_2010}.
In this figure and throughout this work, $f_{\nu}$ represents the
specific flux, the flux per unit frequency $\nu$, in the observer frame.
For these simulations, we use the same input parameters
as \citet{lazzati_and_begelman_2010}: $\Gamma = 1000$,
$k_{B} T_{\gamma} ^{\prime} = \mbox{ 90 eV}$, $N_{\gamma} = 3 \times 10^{6}$,
$N_{e} = 10^{3}$, $N_{\mathrm{collect}} = N_{\gamma} /3$,
mono-energetic electrons initialized to $\gamma_{e} ^{\prime} = 2$,
$\tau_{\mathrm{initial}} =$ 2, 8, 16, and no adiabatic cooling. There is good agreement for
all the simulations.
To quantify this agreement, we performed
Kolmogorov-Smirnov (KS) tests. The probability that our simulations
are drawn from the same distribution as those of
\citet{lazzati_and_begelman_2010} (P-values) are 0.9999 for $\tau_{\mathrm{initial}} = 2$,
0.9862 for $\tau_{\mathrm{initial}} = 8$, and
0.9809 for $\tau_{\mathrm{initial}} = 16$. This good agreement
demonstrates that our MC photospheric code is working properly.

We next perform a test to determine if $10^{3}$ electrons are enough to represent an
electron distribution. If there are not enough electrons, the photon
spectrum will look very noisy. Previous MC photospheric studies
\citep{lazzati_and_begelman_2010,chhotray_and_lazzati_2015} found
that $10^{3}$ electrons are enough to represent an electron distribution;
however, their simulations were performed for lower $N_{\gamma}/N_{e}$.
To determine if $10^{3}$ electrons are enough for our
$N_{\gamma}/N_{e} = 10^{5}$ simulations,  we perform 2
simulations with $N_{\gamma} = 10^{9}$, $N_{e} = 10^{4}$ (10 times
more photons and electrons) and compared them to simulation results with
$N_{\gamma} = 10^{8}$, $N_{e} = 10^{3}$ in right panel of Figure \ref{fig1}
(we used the input parameters described in Section
\ref{input_param_sect} for mono-energetic
electrons with initial $\gamma_{e} ^{\prime}=$ 2, 30
$k_{B}T_{\gamma} ^{\prime} =$ 300 eV, $\Gamma=300$, and adiabatic
cooling). From KS tests,
the probability that the two $\gamma_{e} ^{\prime} = 2$
[$\gamma_{e} ^{\prime} = 30$] simulations are drawn from the same
distribution is 0.9999 [0.9999]. This good agreement
explicitly demonstrates that $10^{3}$ electrons are enough for
$N_{\gamma}/N_{e} = 10^{5}$ simulations.
%
%
%

\section{Parameters Considered for MC Simulations with Seed BB Photons}
\label{paramameter_search_sec}

We now discuss the range of parameters we consider for our
simulations. The main input parameters that affect the output spectrum
are $k_{B} T_{\gamma} ^{\prime}$ (determines the energy of the
majority of the photons), $\gamma_{e} ^{\prime}$ (determines energy in the
electrons), and $\tau_{\mathrm{initial}}$ (determines the average
number of scatterings before a photon arrives at the photosphere).
$L$ determines the physical scales of the system by determining
$R_{\mathrm{initial}}$, $R_{\mathrm{photosphere}}$,
$n_{e} ^{\prime}$, and $\ell_{\mathrm{mfp}} ^{\prime}$, but it does
not affect the number of scatterings or the shape of the spectrum.
The main effect of $\Gamma$ is to Doppler boost the photon spectrum to
the observer frame. In addition, $\Gamma$
also determines the physical scales of the system by determining
$R_{\mathrm{initial}}$, $R_{\mathrm{photosphere}}$,
$n_{e} ^{\prime}$, and $\ell_{\mathrm{mfp}} ^{\prime}$.

For our simulations, we considered $k_{B} T_{\gamma} ^{\prime} =$ 30 eV,
100 eV, 300 eV.
One way the photon temperature can
affect the output spectrum is through the production of
electron-positron pairs by photon annihilation, which would decrease
(increase) the number of photons (electrons) in a simulation.
However, since the typical photon energies we are considering
$k_{B} T_{\gamma} ^{\prime} \sim 30\mbox{ eV}-300 \mbox{ eV}$
are much less than $m_{e} c^{2}$, electron-positron pair production is
expected to be unimportant and we neglect it for our
simulations.
Another more important effect $k_{B} T_{\gamma} ^{\prime}$ has on the simulation
results is on the cooling of electrons. The photons in the
jet-comoving frame are more energetic for larger
$k_{B} T_{\gamma} ^{\prime}$ and more energetic photons will
cool the electrons faster when they undergo multiple scatterings.

$\gamma_{e} ^{\prime}$ is an important parameter since it determines
the available energy electrons have to transfer to photons.
The smallest value we consider for $\gamma_{e} ^{\prime}$ is 2.
In the photospheric process, in order to avoid synchrotron cooling
from the magnetic field that is expected to be present in the jet,
the synchrotron emission is taken to be self-absorbed.
The largest $\gamma_{e} ^{\prime}$ that can be considered
is found by setting the optical depth for
synchrotron self-absorption equal to 1. Below, we calculate the largest
$\gamma_{e} ^{\prime}$ allowed for MB and PL distributions of
electrons.

For a MB distribution distribution of electrons, the
synchrotron self-absorption optical depth $\tau_{\mathrm{syn}} ^{\mathrm{MB}}$
is given by
\citep[][]{rybicki_and_lightman_1979,lazzati_and_begelman_2010}
\begin{equation}
\tau_{\mathrm{syn}} ^{\mathrm{MB}} = \frac{10^{6}}{(\gamma_{e} ^{\prime}) ^{5}
\epsilon_{B} ^{1/2} (E _{ \gamma} ^{\prime} / \mbox{1keV}) ^{2}} .
\label{opt_depth_sync_SA}
\end{equation}
In this equation, $\epsilon_{B} = U_{B}/U_{\mathrm{rad}}$,
where $U_{B}$ ($U_{\mathrm{rad}}$) is the energy density in the
magnetic field (radiation) and we note that
$\tau_{\mathrm{syn}} ^{\mathrm{MB}}$ depends on the energy of the
photons in the comoving frame.
Setting $\tau_{\mathrm{syn}} ^{\mathrm{MB}} = 1$, for $\epsilon_{B} \sim 0.1$
(magnetic field subdominant to radiation) and
$E _{ \gamma} ^{\prime} \sim \mbox{300 eV} $
($E _{ \gamma} ^{\prime} \sim \mbox{100 eV} $)
[$E _{ \gamma} ^{\prime} \sim \mbox{30 eV} $], the
upper limit we find is $\gamma_{e} ^{\prime} \sim 30$
($\gamma_{e} ^{\prime} \sim 50$)
[$\gamma_{e} ^{\prime} \sim 80$].

For a power-law distribution distribution of electrons, the
synchrotron self-absorption optical depth $\tau_{\mathrm{syn}} ^{\mathrm{PL}}$
is given by \citep{wu_et_al_2003,gou_et_al_2007}
\begin{eqnarray}
\tau_{\mathrm{syn}} ^{\mathrm{PL}} &=& \frac{e}{B ^{\prime}}
\left( \frac{(p-1)(p+2)}{(\gamma_{1} ^{\prime}) ^{-(p-1)} -
(\gamma_{2} ^{\prime}) ^{-(p-1)}} \right) \times \nonumber \\
&&\frac{2 \sqrt{6} \pi^{3/2}}{9} (\gamma_{2} ^{\prime}) ^{-(p+4)}
\mbox{exp}(-1) \sigma_{T} ^{-1} .
\end{eqnarray}
In this equation, $\gamma_{1} ^{\prime}$ ($\gamma_{2} ^{\prime}$) is
the electron Lorentz factor where the power-law begins (ends) and
$B^{\prime}$ is the magnetic field in the jet-comoving frame. From the
expressions $U_{B} = (B^{\prime})^{2}/(8 \pi) $ and
$U_{\mathrm{rad}} = a_{\mathrm{rad}} (T_{\gamma} ^{\prime})^{4}$,
$B^{\prime} = (8 \pi \epsilon_{B} a_{\mathrm{rad}})^{1/2} (T_{\gamma} ^{\prime})^{2}$.
Setting $\tau_{\mathrm{syn}} ^{\mathrm{PL}} =1$, taking
$\gamma_{1} ^{\prime}= 2$, $p=2.4$, and $B ^{\prime}$ corresponding to
$\epsilon_{B} = 0.1$ and $k_{B} T_{\gamma} ^{\prime} = \mbox{ 300 eV}$
($k_{B} T_{\gamma} ^{\prime} = \mbox{ 100 eV}$)
[$k_{B} T_{\gamma} ^{\prime} = \mbox{ 30 eV}$], the upper limit we found for
$\gamma_{2} ^{\prime}$ is $\gamma_{2} ^{\prime} \sim 30$
($\gamma_{2} ^{\prime} \sim 50$)
[$\gamma_{2} ^{\prime} \sim 80$].
Thus, for both a MB distribution and a PL distribution, the maximum
$\gamma_{e} ^{\prime}$ we can consider
for $k_{B} T_{\gamma} ^{\prime} = \mbox{ 300 eV}$
($k_{B} T_{\gamma} ^{\prime} = \mbox{ 100 eV}$)
[$k_{B} T_{\gamma} ^{\prime} = \mbox{ 30 eV}$] is
$\sim 30$ ($\sim 50$) [$\sim 80$]. A summary of the values we
considered for $\gamma_{e} ^{\prime}$ for each value of $k_{B}
T_{\gamma} ^{\prime}$ is given in Table \ref{table1}.
\begin{table*}
\centering
\begin{tabular}{ c c c c}
     $k_{B} T_{\gamma} ^{\prime}$ & Mono-Energetic & MB & PL Distribution \\ \hline
     30 eV & $\gamma_{e} ^{\prime} =$ 2, 30, 80 &
    $\frac{k_{B} T_{e} ^{\prime}}{(\gamma_{\mathrm{ad, el}} -1)m_{e} c^{2}} = $ (2-1), (80-1) &
    $\gamma_{e,1} ^{\prime}=$ 2,
    $\gamma_{e,2} ^{\prime}=$ 80,
    $p=2.4$ \\ \hline
    100 eV & $\gamma_{e} ^{\prime} =$ 2, 30, 50 &
    $\frac{k_{B} T_{e} ^{\prime}}{(\gamma_{\mathrm{ad, el}} -1)m_{e} c^{2}} = $ (2-1), (50-1) &
    $\gamma_{e,1} ^{\prime}=$ 2,
    $\gamma_{e,2} ^{\prime}=$ 50,
    $p=2.4$ \\ \hline
   300 eV & $\gamma_{e} ^{\prime} =$ 2, 10, 30 &
    $\frac{k_{B} T_{e} ^{\prime}}{(\gamma_{\mathrm{ad, el}} -1)m_{e} c^{2}} = $ (2-1), (30-1) &
    $\gamma_{e,1} ^{\prime}=$ 2,
    $\gamma_{e,2} ^{\prime}=$ 30,
    $p=2.4$ \\ \hline
\end{tabular}
\caption{$\gamma_{e} ^{\prime}$ values we consider for our
  simulations for each value of $k_{B} T_{\gamma} ^{\prime}$ and the
  3 different electron distributions we consider. For the MB
  distribution, we give the value of
  $k_{B} T_{e} ^{\prime} = (\gamma_{\mathrm{ad, el}} -1)(\gamma_{e} ^{\prime} -1) m_{e} c^{2}$,
  where $\gamma_{\mathrm{ad, el}} = (4 \gamma_{e} ^{\prime} + 1)/(3 \gamma_{e} ^{\prime})$
  is the electron adiabatic index, since $k_{B} T_{e} ^{\prime}$ measures the kinetic energy of
  the electrons. \label{table1}}
\end{table*}

$\tau_{\mathrm{initial}}$ affects the shape of the spectrum since it
determines the average number of scatterings a photon experiences
before escaping the photosphere.
The average number of scatterings for a photon is
$\sim 2 \tau_{\mathrm{initial}}$ \citep{begue_et_al_2013}, not
$\sim (\tau_{\mathrm{initial}})^{2}$, since as the GRB relativistic outflow moves
outward, $n_{e} ^{\prime}$ decreases as $R^{2}$
(Equation \ref{electron_density_comoving}). For our simulations, we
considered $\tau_{\mathrm{initial}} = $ 2, 8, 16.

Lastly, since $N_{\gamma} \gg N_{e}$ for our simulations, the electrons
rapidly cool by IC scatterings and then the electrons no longer have much
energy to transfer to the photons. Thus, we also
considered electron re-heating. At a given
$\tau_{\mathrm{initial}}$, the total number of scatterings expected
is $\sim 2 \tau_{\mathrm{initial}} N_{\gamma}$. To re-heat the
electrons, we first specify the number of re-heating events
$N_{\mathrm{rh}}$ we choose to consider. Then, after
$(2 \tau_{\mathrm{initial}} N_{\gamma})/(N_{\mathrm{rh}} + 1)$
scatterings, we re-heat the electrons to the same distribution that
they were initialized to (we divide by $N_{\mathrm{rh}} + 1$
since the total number of heating events is the initial heating event
plus $N_{\mathrm{rh}}$ events). We adopt this methodology to have
the re-heating events evenly spaced within the total number of
scatterings.

In this electron re-heating scheme, since all the electrons
  are re-accelerated, it corresponds to a global heating mechanism. In
GRBs, the heating mechanism is likely to be global, i.e. occurring
throughout the causally connected part of the jet, since observations
show that the conversion of jet energy to gamma-ray radiation is an
efficient process
\citep{granot_et_al_2006,fan_and_piran_2006,zhang_et_al_2007}. 
The electrons can be re-accelerated by magnetic
reconnection in a Poynting flux dominated jet or by shocks in a baryon
dominated jet.
%
%
%

\section{Simulation Results for Comptonization of Seed BB Spectrum}
\label{results_and_interpretation_sect}

In this section, we first show our results for one dissipation event,
where electrons are only accelerated once at the start of the simulation. Then, we
consider electron re-heating, where the electrons are re-accelerated
back to their initial distribution $N_{\mathrm{rh}}$ times.
\subsection{Simulation Results for One Dissipation Event}

\begin{figure*}
\begin{center}$
\begin{array}{l}
\begin{array}{ll}
\hspace{-10mm}
\includegraphics[scale=0.45]{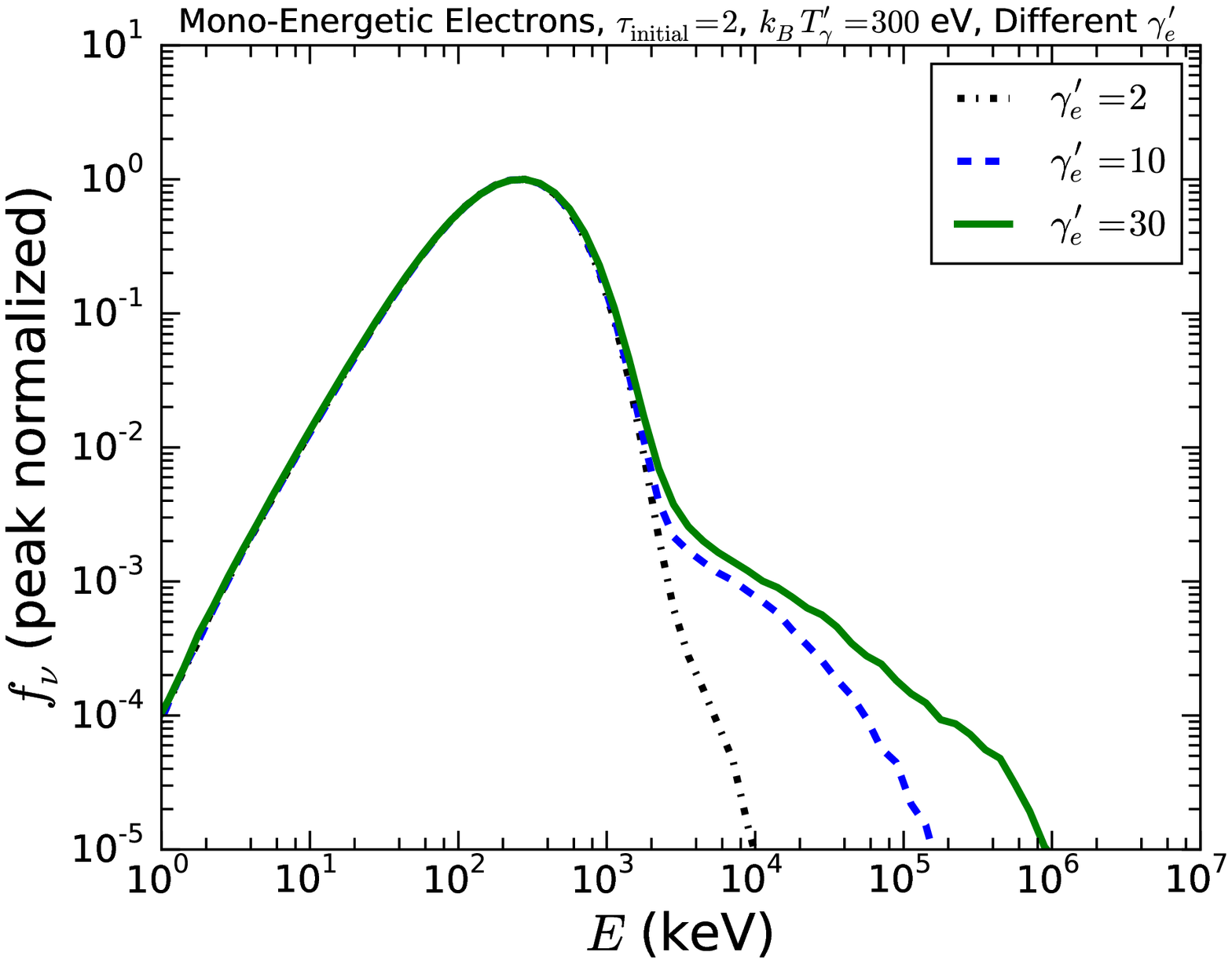} &
\hspace{0mm}
\includegraphics[scale=0.45]{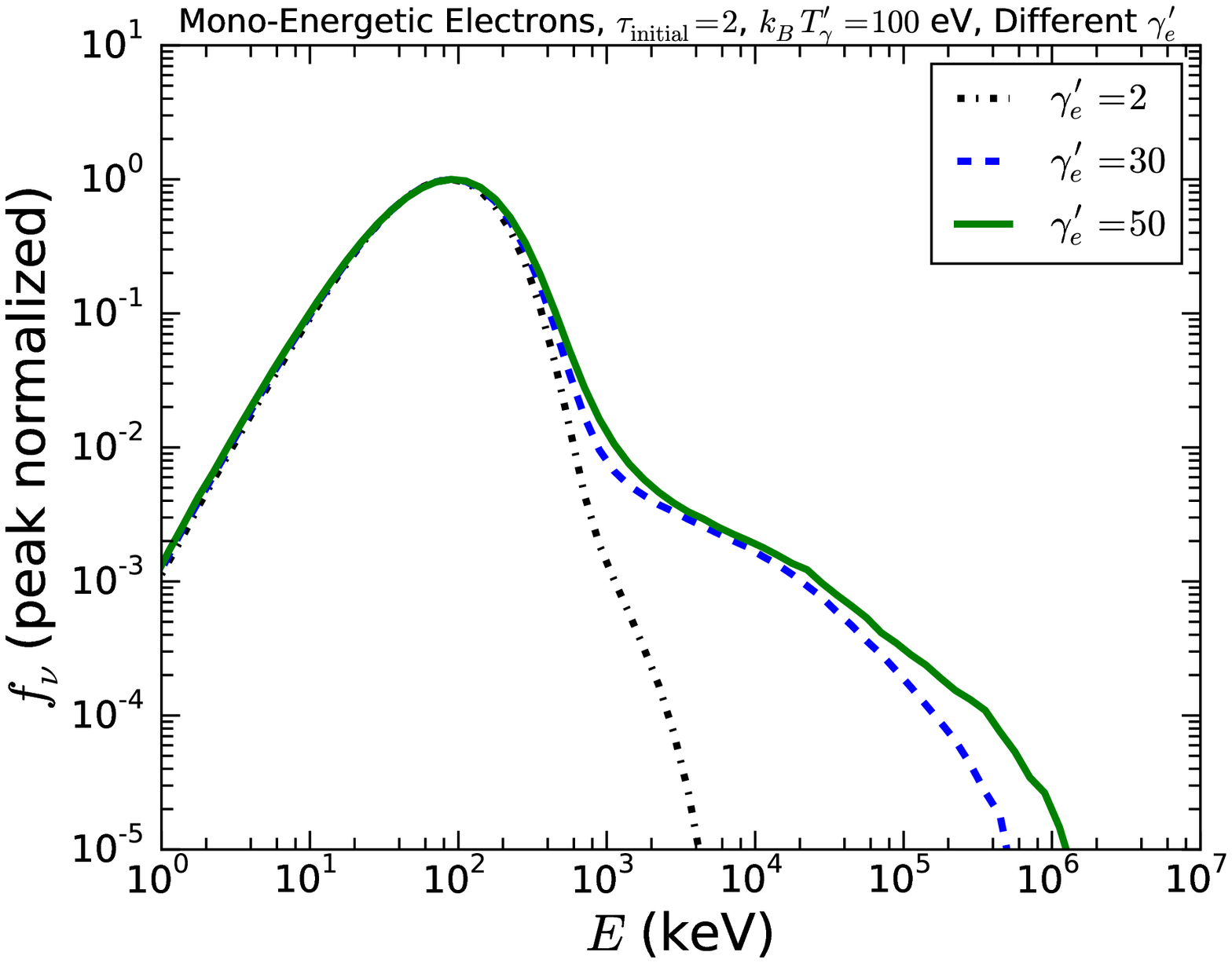}
\end{array}
\\
\hspace{-10mm}
\includegraphics[scale=0.45]{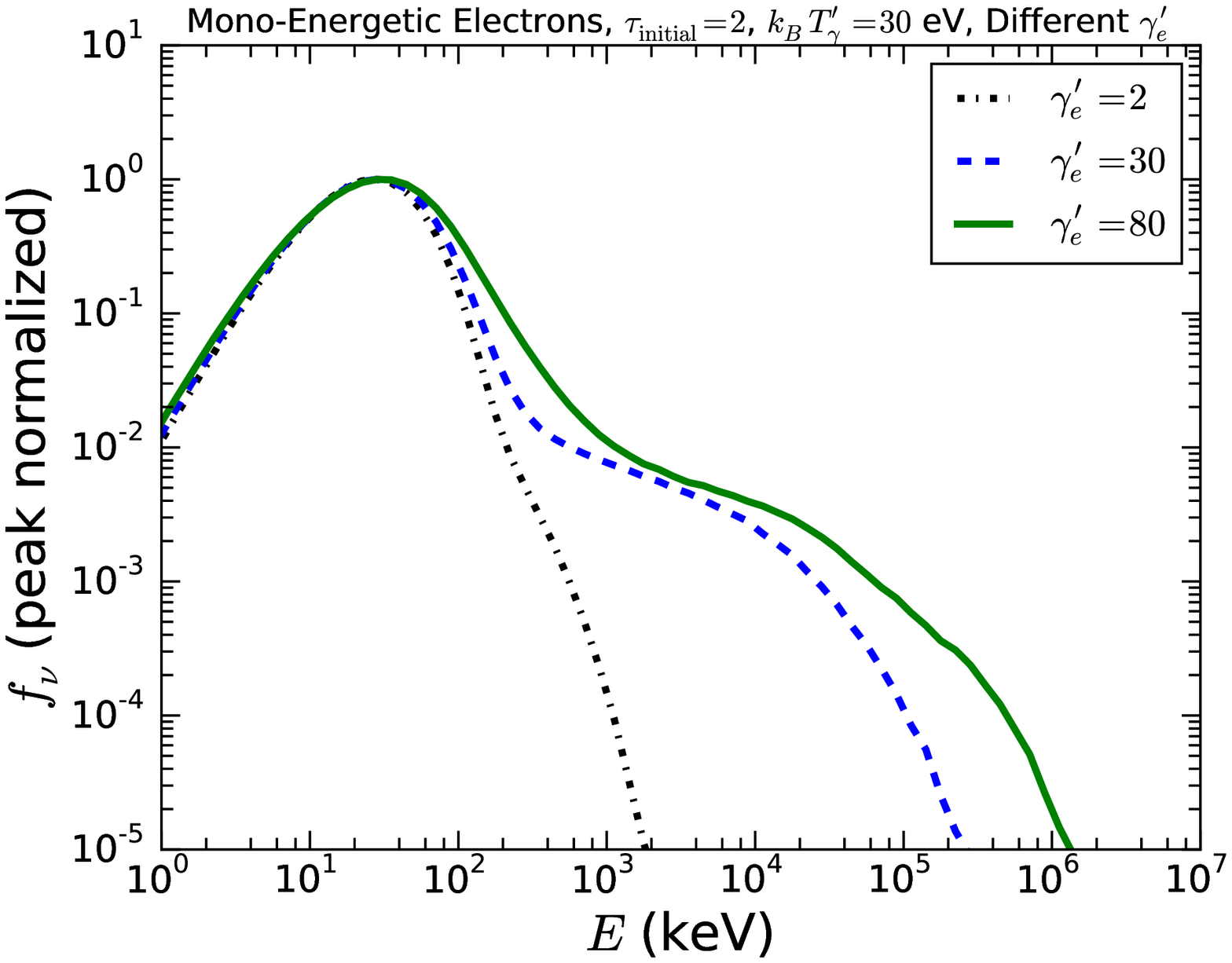}
\end{array}$
\end{center}
\caption{\textit{Top-Left panel:} Simulation results for the Comptonization of seed BB photons with
  $k_{B} T_{\gamma} ^{\prime} = 300 \mbox{ eV}$, $\Gamma=300$ with
  mono-energetic electrons with initial
  $\gamma_{e} ^{\prime} = $ 2, 10, 30 and $\tau_{\mathrm{initial}} =2$.
  \textit{Top-Right panel:} Same as \textit{Top-Left panel} for
  $k_{B} T_{\gamma} ^{\prime} = 100 \mbox{ eV}$, $\Gamma=300$ and mono energetic electrons with
  initial $\gamma_{e} ^{\prime} = $ 2, 30, 50.
  \textit{Bottom-Left panel:} Same as \textit{Top-Left panel}
  for $k_{B} T_{\gamma} ^{\prime} =
  30 \mbox{ eV}$, $\Gamma=300$ and mono energetic electrons with
  initial $\gamma_{e} ^{\prime} = $ 2, 30, 80. \label{fig2}}
\end{figure*}

\begin{figure*}
\begin{center}$
\begin{array}{l}
\begin{array}{ll}
\hspace{-10mm}
\includegraphics[scale=0.45]{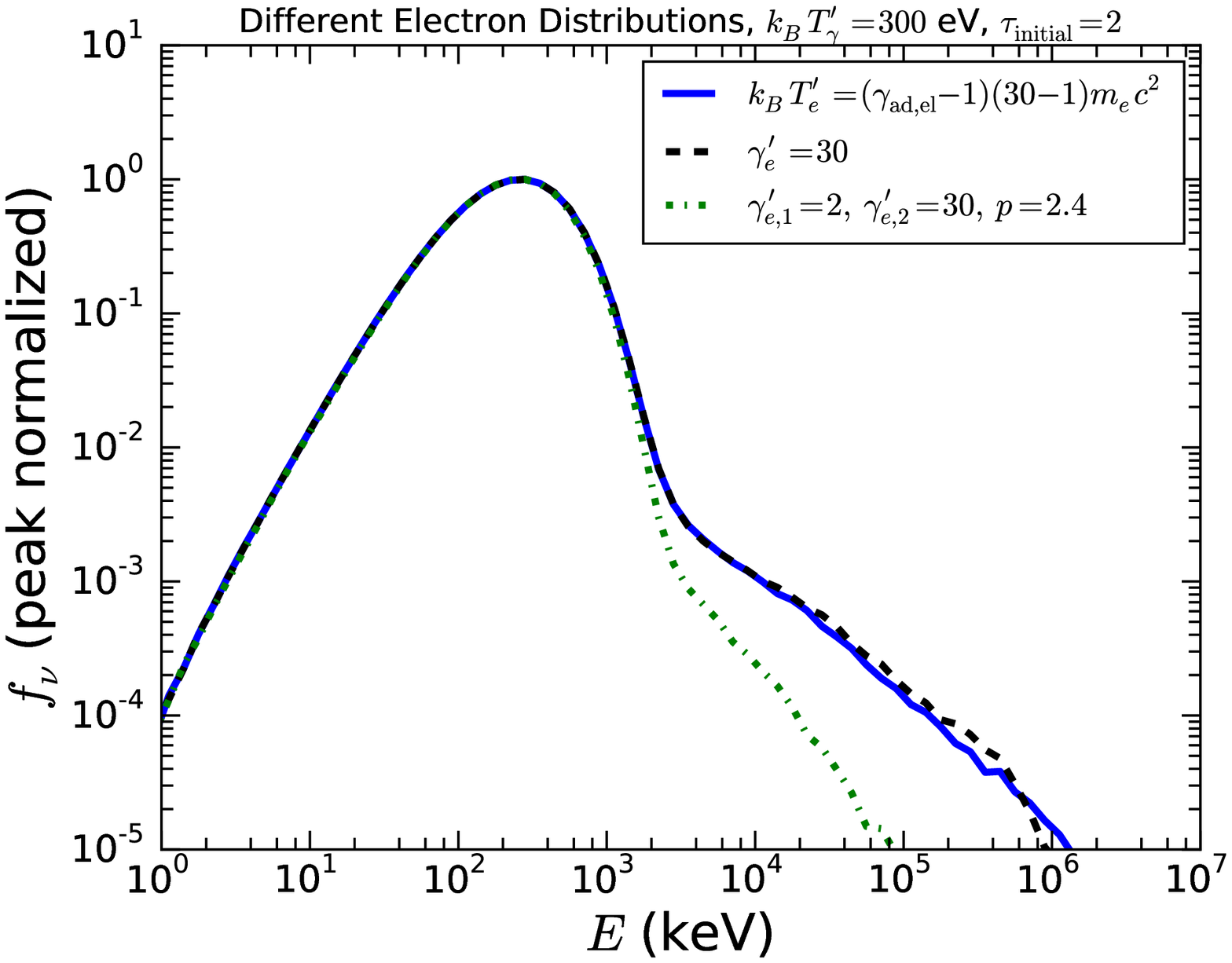} &
\hspace{0mm}
\includegraphics[scale=0.45]{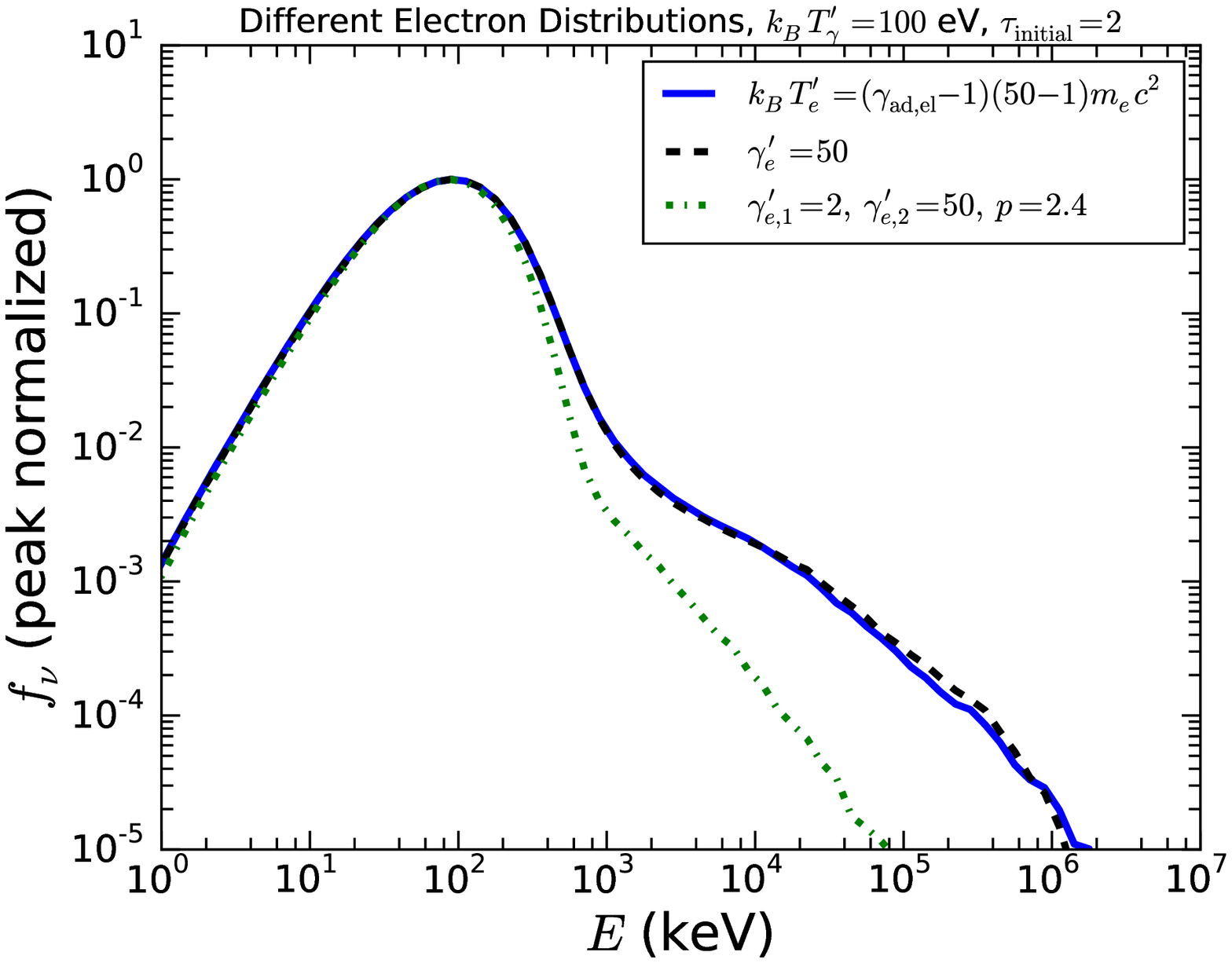}
\end{array}
\\
\hspace{-10mm}
\includegraphics[scale=0.45]{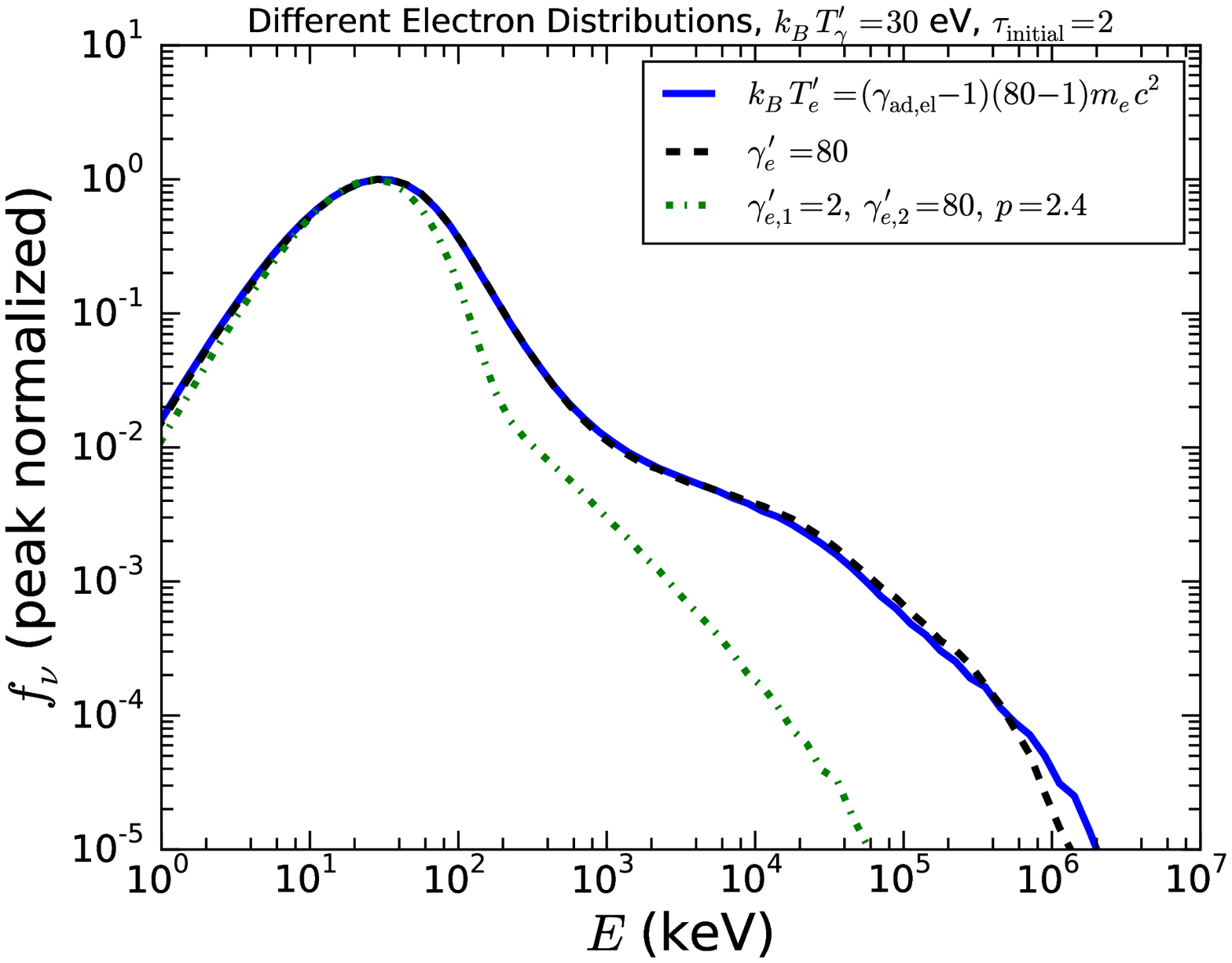}
\end{array}$
\end{center}
\caption{\textit{Top-Left panel:} Simulation results for the Comptonization of seed BB photons
  with $k_{B} T_{\gamma} ^{\prime} = 300 \mbox{ eV}$, $\Gamma=300$
  with electrons following MB, mono energetic, and PL distributions,
  and $\tau_{\mathrm{initial}} = 2$. For each distribution, we
  considered the largest value of $\gamma_{e} ^{\prime}$ we can
  consider for $k_{B} T_{\gamma} ^{\prime} = \mbox{ 300 eV}$
  (see discussion in Section \ref{paramameter_search_sec} and Table \ref{table1}).
  \textit{Top-Right panel:} Same as \textit{Top-Left panel}  but with
  $k_{B} T_{\gamma} ^{\prime} = 100 \mbox{ eV}$, $\Gamma=300$.
  \textit{Bottom-Left panel:} Same as as \textit{Top-Left panel} but with
  $k_{B} T_{\gamma} ^{\prime} = 30 \mbox{ eV}$, $\Gamma=300$.
\label{fig3}}
\end{figure*}

\begin{figure*}
\begin{center}$
\begin{array}{l}
\begin{array}{ll}
\hspace{-10mm}
\includegraphics[scale=0.45]{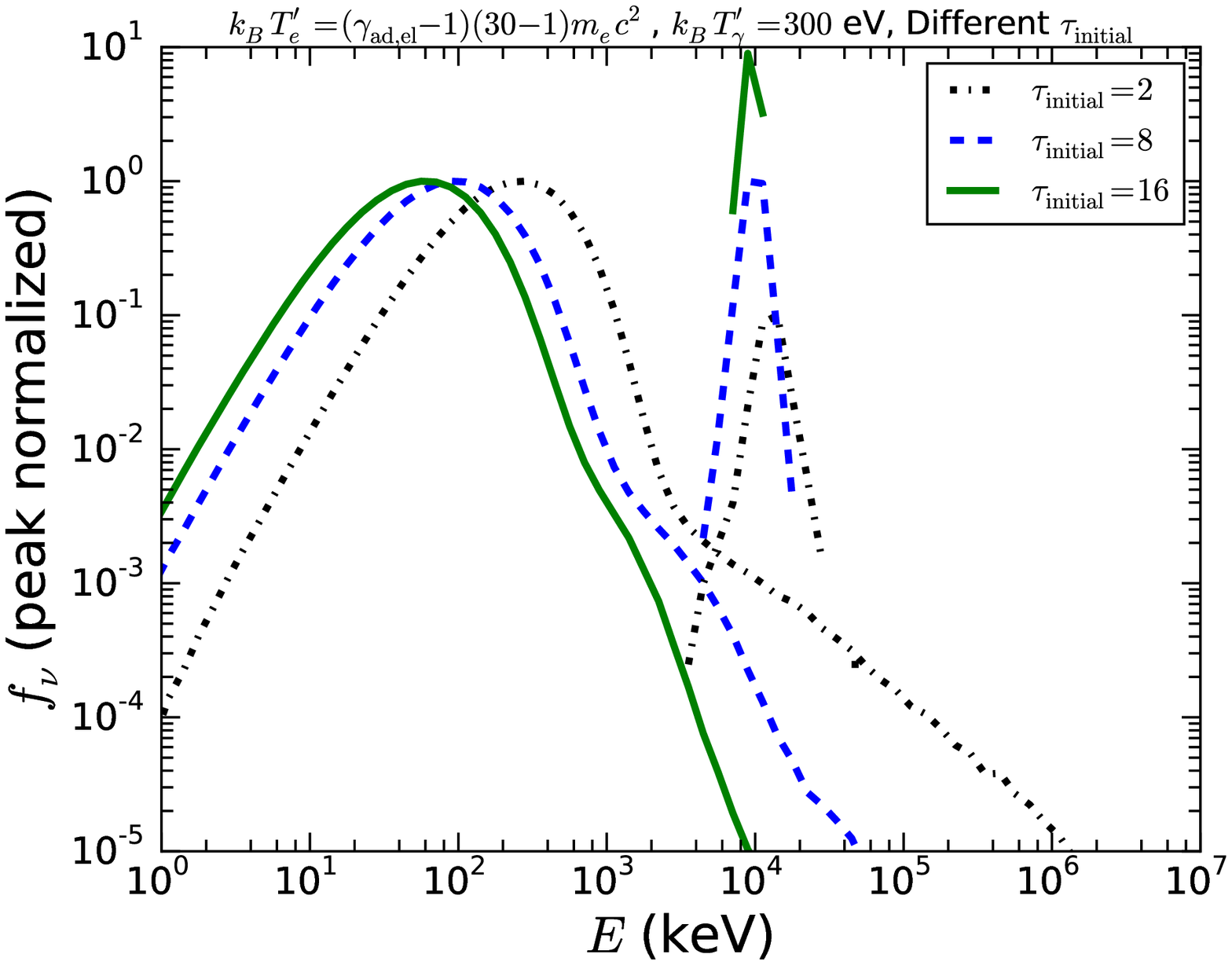} &
\hspace{0mm}
\includegraphics[scale=0.45]{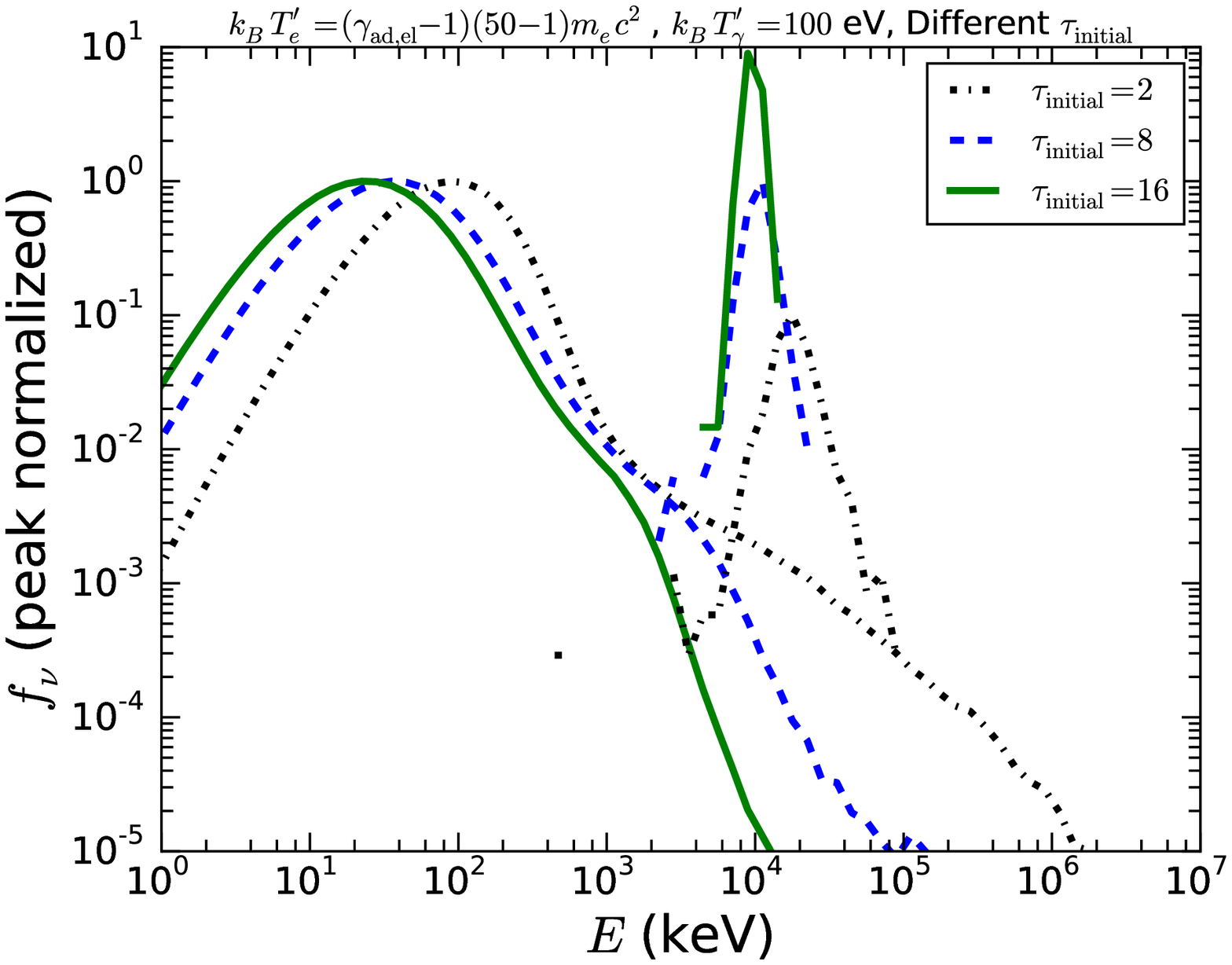}
\end{array}
\\
\hspace{-10mm}
\includegraphics[scale=0.45]{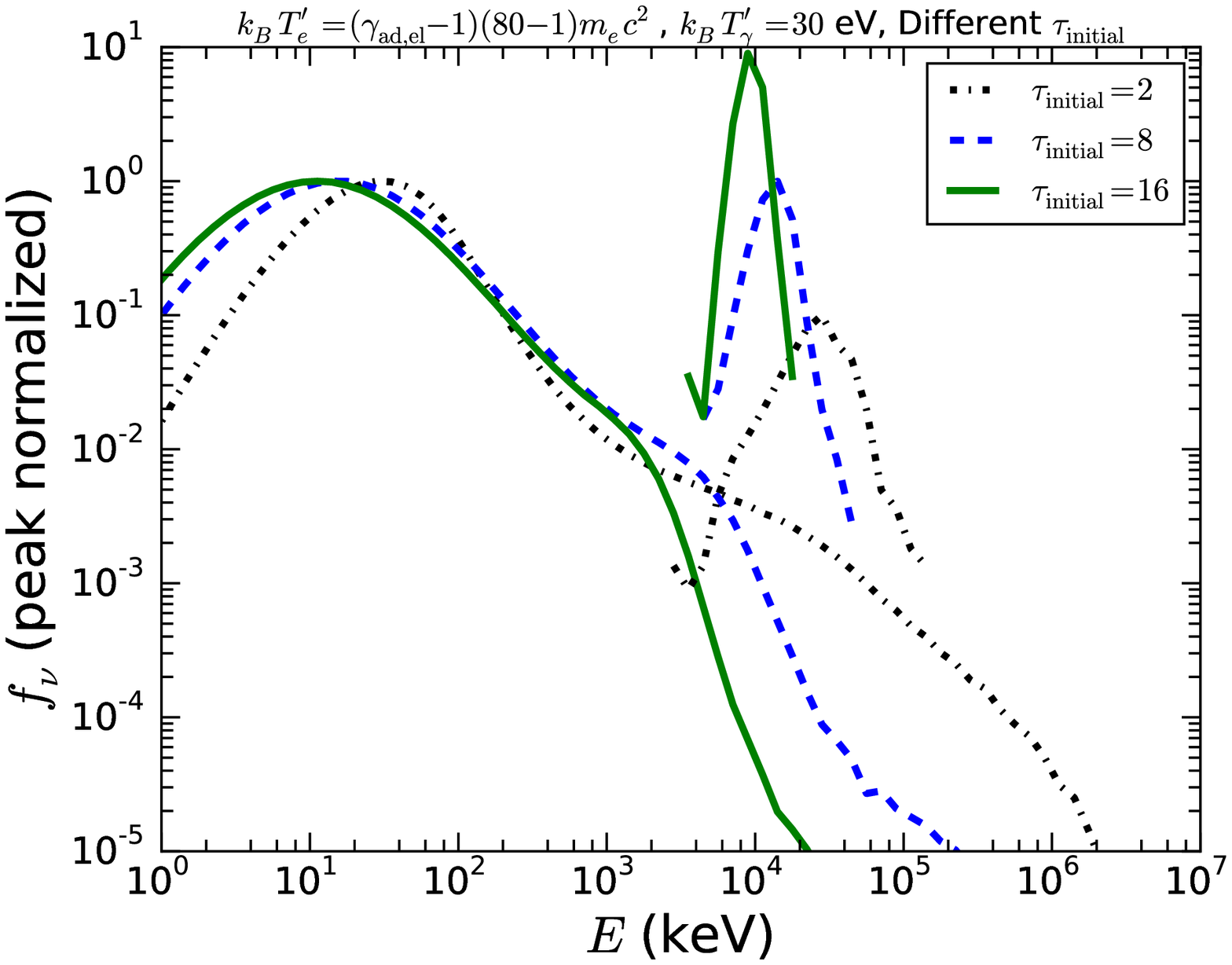}
\end{array}$
\end{center}
\caption{\textit{Top-Left panel:} Simulation results for the Comptonization of seed BB photons with
  $k_{B} T_{\gamma} ^{\prime} = 300$ eV, $\Gamma=300$ with
  MB electrons with initial
  $\gamma_{e} ^{\prime} \sim $ 30 and $\tau_{\mathrm{initial}} = $ 2, 8 16.
  The photon spectra are peak normalized and the electron spectra 
  (spectra in the top right) are shifted by a factor of
  10 for each $\tau_{\mathrm{initial}}$ to better see if there is any
  change as $\tau_{\mathrm{initial}}$ becomes larger. For both the 
  photons and the electrons, we are plotting their
  energy spectrum $f_{\nu} = E N_{E}$.
  \textit{Top-Right panel:}
  Same as \textit{Top-Left panel}, but with
  $k_{B} T_{\gamma} ^{\prime} = 100 \mbox{ eV}$, $\Gamma=300$
  and MB electrons with initial $\gamma_{e} ^{\prime} \sim
  50$. \textit{Bottom-Left panel:} Same as \textit{Top-Left panel}, but with
  $k_{B} T_{\gamma} ^{\prime} = 30 \mbox{ eV}$, $\Gamma=300$
  and MB electrons with initial $\gamma_{e} ^{\prime} \sim  80$. 
 The $\tau_{\mathrm{initial}} =$ 8, 16 simulations in 
  the \textit{Bottom-Left panel} display a power-law with a spectrum
  $f_{\nu} \propto \nu^{-1.2}$.
\label{fig4}}
\end{figure*}

In Figure \ref{fig2}, we show simulations for
$\tau_{\mathrm{initial}} = 2$, the three values
we considered for $k_{B} T_{\gamma} ^{\prime}$,
and mono-energetic electrons with the values of
$\gamma_{e} ^{\prime}$ shown in Table \ref{table1}.
As the reader many have noticed in Figure \ref{fig1},
the most striking feature of the output spectrum is the sharp drop in $f_{\nu}$
by $\sim 2$ orders of magnitude after the peak energy,
$E_{\mathrm{pk}}$. Unlike previous
studies for $N_{\gamma}/N_{e} = 1 -10^{4}$, our results for
$N_{\gamma}/N_{e} = 10^{5}$ do not show a power-law immediately after
the peak energy. After the drop in
$f_{\nu}$, the simulations with $\gamma_{e} ^{\prime} = 2$
in the three panels continue to
decline rapidly. On the other hand, the simulations with
 $\gamma_{e} ^{\prime} \sim 10-80$
in each panel display a power-law for
$\sim 2-3$ decades in energy before declining rapidly again.
The highest energy the photons
near the BB peak can attain after one scattering
is $ \sim 4 \times (100 \mbox{ eV}) \Gamma (\gamma_{e} ^{\prime})^{2}$
$\sim 10^{5} \mbox{ keV}$ for
$\gamma_{e} ^{\prime} \sim 50$. Once the photons reach these
energies, IC scattering is highly Klein-Nishina (KN) suppressed, leading to an
exponential decay in the spectrum for higher energies.
In Figure \ref{fig3}, we show simulation results for
$\tau_{\mathrm{initial}} = 2$, the three values
we considered for $k_{B} T_{\gamma} ^{\prime}$,
and the largest $\gamma_{e} ^{\prime}$ we considered for each
distribution in Table \ref{table1}.
In each panel, the PL distribution simulations display the
least broadened spectrum. This is due to the fact that the PL
distribution contains the least energetic electrons among these three
distributions. The mono-energetic and MB
electron distribution simulations show similar results since the two
distributions are similar; the MB distribution has an average
$\gamma_{e} ^{\prime}$ very close to the $\gamma_{e} ^{\prime}$ value we consider
for the mono-energetic electrons. After the sharp drop in $f_{\nu}$ in
each panel, all the simulations
display a single power-law spectrum for
$\sim 3$ decades in energy.
In Figure \ref{fig4}, we show simulation results for
$\tau_{\mathrm{initial}} =$ 2, 8, 16, the three values of
$k_{B} T_{\gamma} ^{\prime}$ we considered, and MB
electrons with the largest value we considered for
$\gamma_{e} ^{\prime}$ (see Table \ref{table1}).
In addition, in each panel,
we also plot the energy
spectrum $f_{\nu} = E N_{E}$ of the kinetic energies of the electrons
at the end of each simulation in the observer frame, i.e.
$m_{e} c^{2} (\gamma_{e}-1) \Gamma$. Increasing
$\tau_{\mathrm{initial}}$ has two effects on the output
spectrum: decreasing the number of photons at higher energies and
increasing the broadening of the spectrum. For larger $\tau_{\mathrm{initial}}$, the additional
scatterings allow for the high-energy photons to transfer energy back
to the electrons. The increase in broadening of the spectrum from
$k_{B} T_{\gamma} ^{\prime} = 300 \mbox{ eV}$ (Top-Left Panel) to
$k_{B} T_{\gamma} ^{\prime} = 30 \mbox{ eV}$ (Bottom-Left Panel) is
due to a couple of effects:
1. the larger $\gamma_{e} ^{\prime}$ considered for
$k_{B} T_{\gamma} ^{\prime} = 30 \mbox{ eV}$ allows for the electrons
to transfer more energy to the photons. 2. for
$k_{B} T_{\gamma} ^{\prime} = 30 \mbox{ eV}$, the photons have lower
energy in the jet-comoving frame, and thus, cool the electrons more
slowly, allowing for more photons to be upscattered to higher
energies. In each panel, the simulations
with larger $\tau_{\mathrm{initial}}$ have a lower peak-energy due to the
adiabatic cooling of photons. In the top-left panel, the three
simulations show a sharp drop in $f_{\nu}$ by $\sim 2$ orders of
magnitude above the peak energy. In the top-right panel, the
$\tau_{\mathrm{initial}} =$ 8, 16 almost show a power-law above the
peak-energy, but the spectrum still declines rapidly. Fitting a power-law
to the  $\tau_{\mathrm{initial}} =$ 8, 16 simulations, we find
a steep spectrum $f_{\nu} \propto \nu^{-2}$.
In the bottom-left panel, the $\tau_{\mathrm{initial}} =$ 8, 16
simulations show a power-law above the peak-energy, with a
$f_{\nu} \propto \nu^{-1.2}$ spectrum, in agreement with the Band function.
However, the peak-energy for these two spectra is 
$\sim \mbox{30 keV}$, 10 times smaller than the typical peak-energy of
the prompt emission.

The electron distributions at the end of all the simulations
display a Maxwell-Boltzmann
distribution. In each panel, as $\tau_{\mathrm{initial}}$
increases, the electron distributions become narrower and the electron
temperature decreases. This is due to the fact that the
additional scatterings allow for the electrons to transfer more energy
to the photons.

In summary, we searched a wide parameter space for the photospheric
process in Figures \ref{fig2}-\ref{fig4}. In  Figure \ref{fig2},
for $\tau_{\mathrm{initial}} = 2$, there is a sharp drop in $f_{\nu}$
above the peak-energy by $\sim 2$ orders of magnitude for all the
photon temperatures and $\gamma_{e} ^{\prime}$ values we considered.
In Figure \ref{fig3}, we determined that the electron distribution
does not have a large impact on the simulation results. In
Figure \ref{fig4}, we determined that considering a larger optical
depth and a lower photon temperature broadens the BB spectrum more, with
a power-law spectrum being developed above the peak-energy for the
lowest photon temperature we considered
($k_{B} T_{\gamma} ^{\prime} = \mbox{ 30 eV}$).
However, the peak-energy for 
the $k_{B} T_{\gamma} ^{\prime} = \mbox{ 30 eV}$ 
and $\tau_{\mathrm{initial}} =$ 8, 16 simulations is a factor of 10
smaller than the typical peak-energy of the prompt emission.
In the next subsection,
we consider electron re-heating to determine
the additional energy that needs to be added to the
electrons of the $k_{B} T_{\gamma} ^{\prime} = \mbox{ 100 eV}$,
300 eV simulations to produce a power-law above the
peak-energy.
\subsection{Simulation Results with Electron Re-heating}

\begin{figure*}
\begin{center}$
\begin{array}{l}
\begin{array}{ll}
\hspace{-10mm}
\includegraphics[scale=0.45]{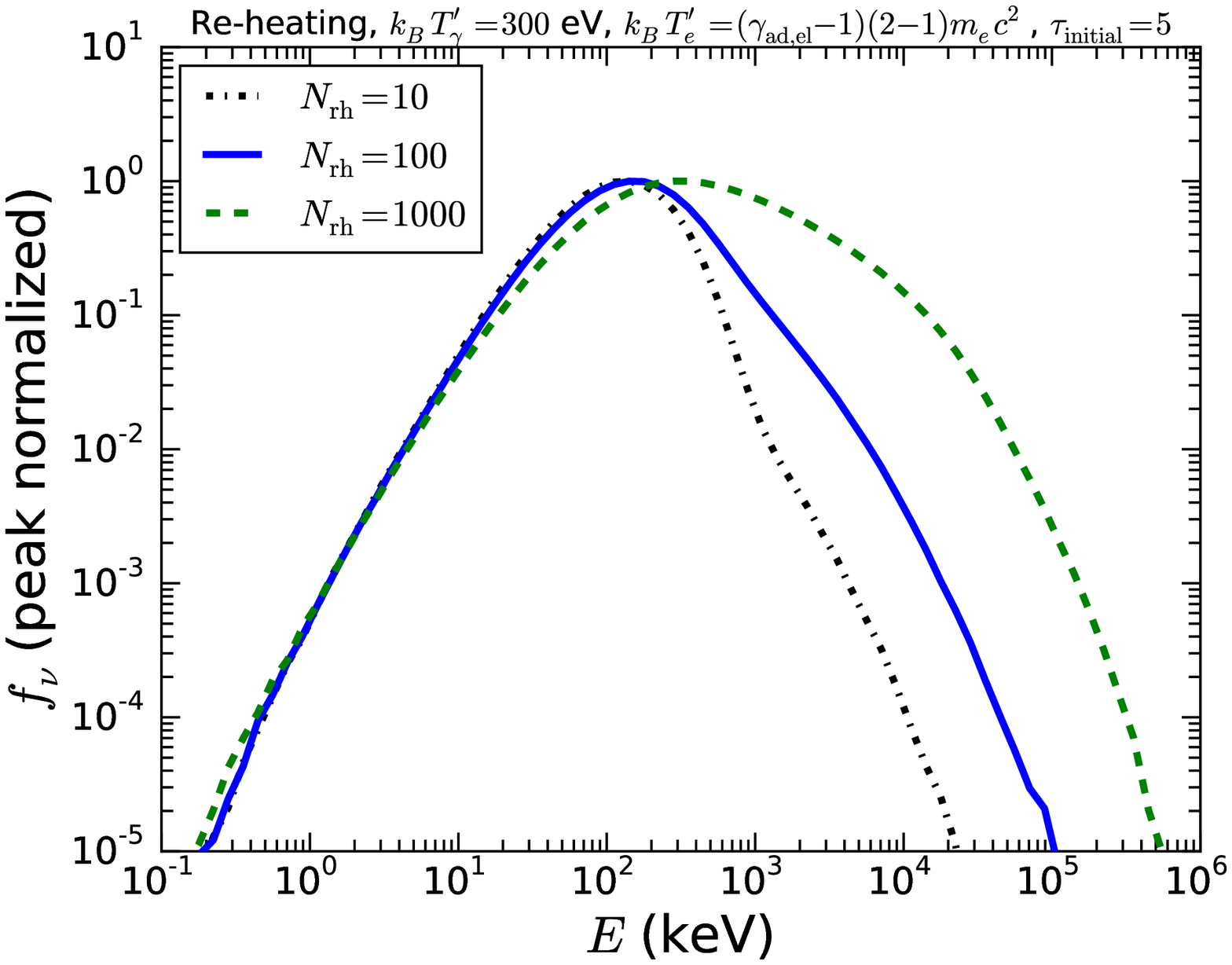} &
\hspace{0mm}
\includegraphics[scale=0.45]{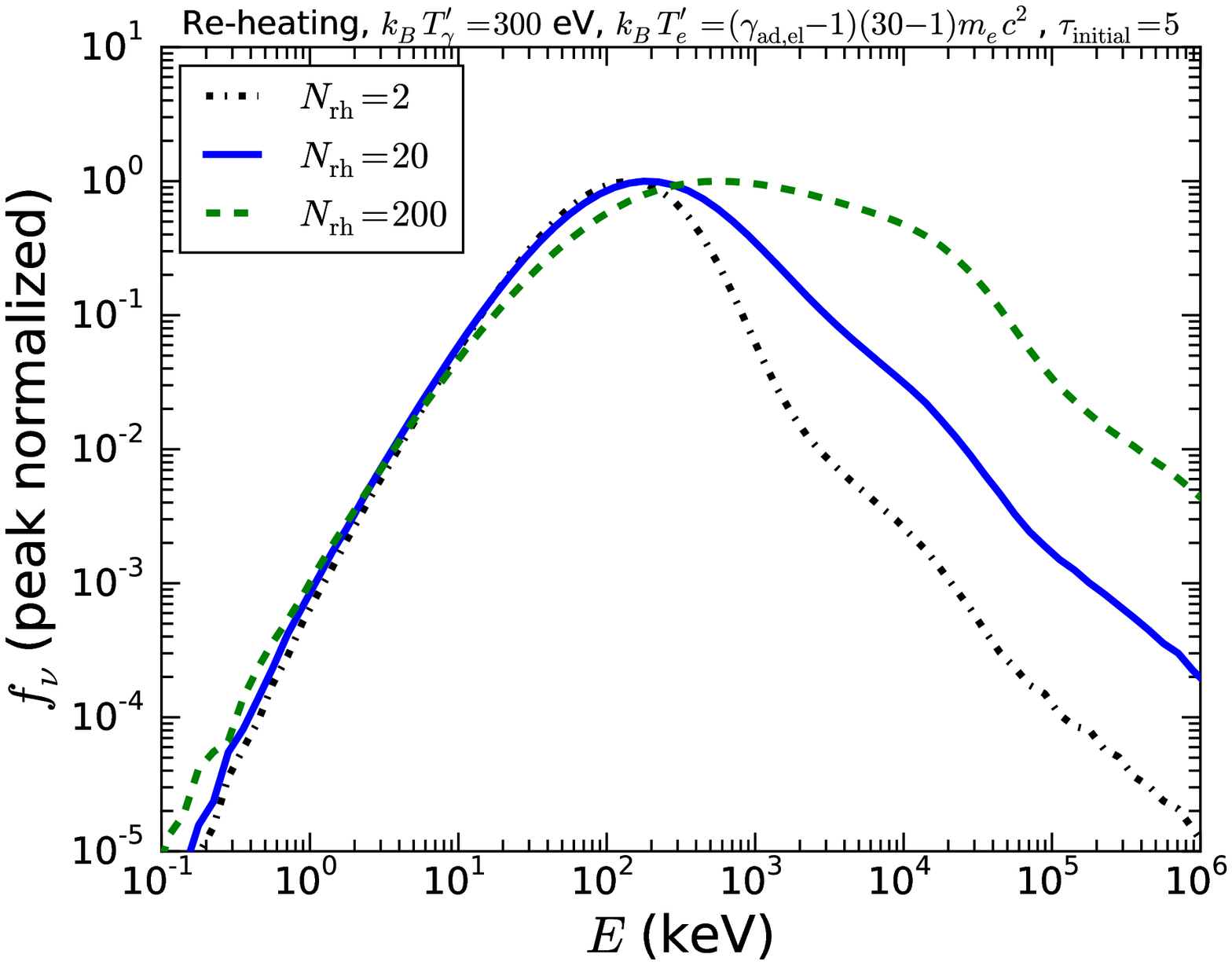}
\\
\hspace{-10mm}
\includegraphics[scale=0.45]{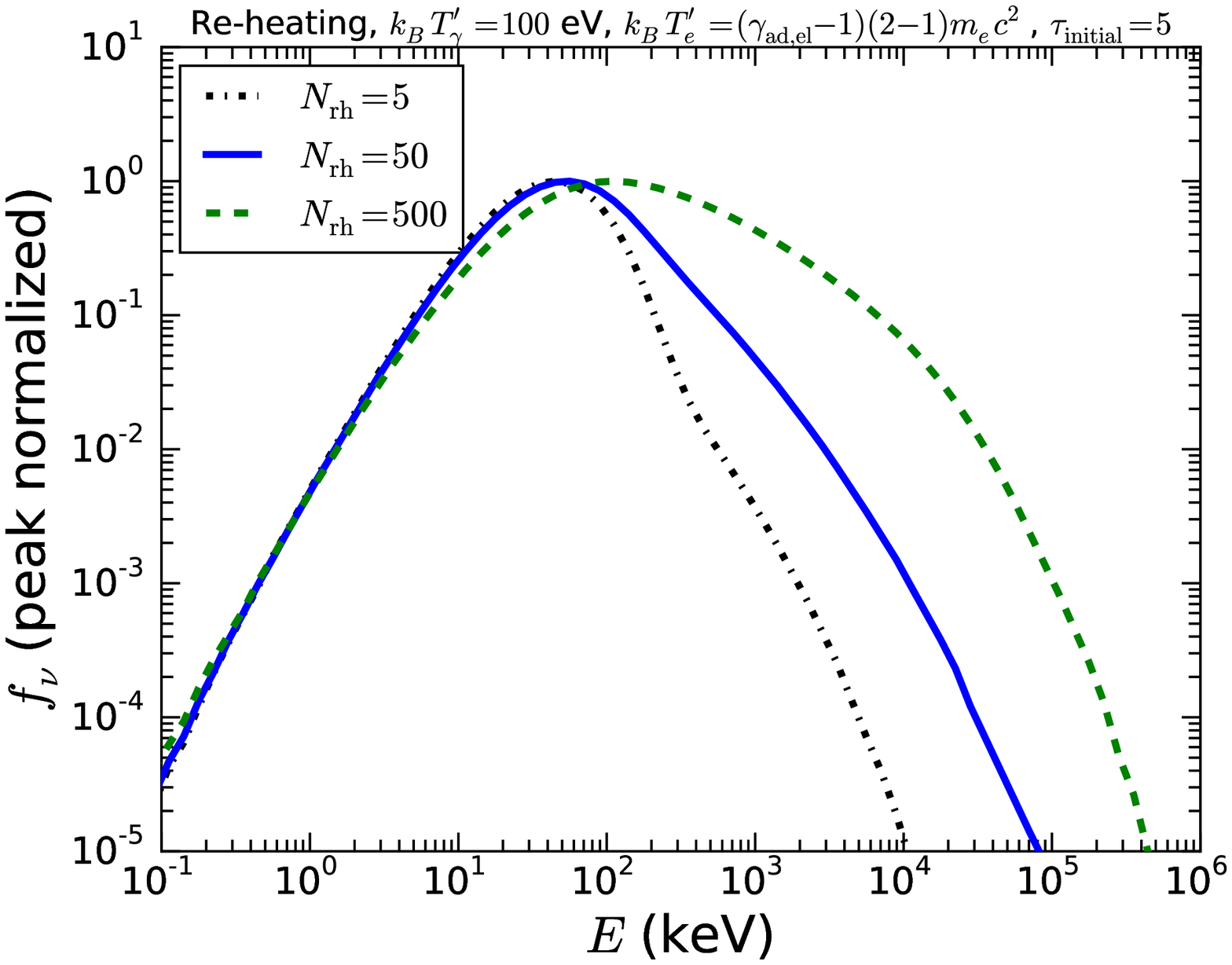} &
\hspace{0mm}
\includegraphics[scale=0.45]{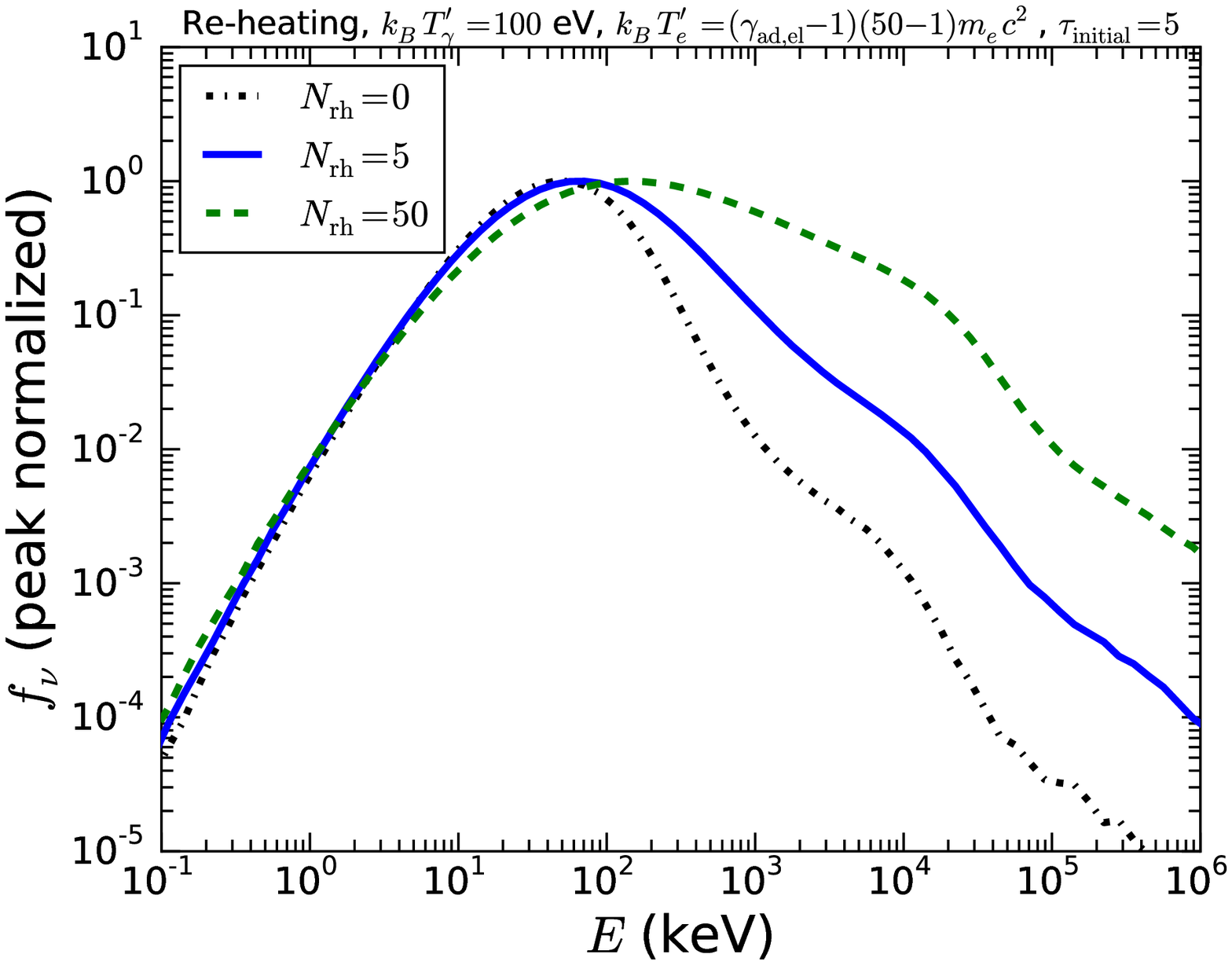}
\end{array}
\end{array}$
\end{center}
\caption{\textit{Top-left Panel:} Simulation results for the Comptonization of seed BB photons
  with $k_{B} T_{\gamma} ^{\prime} = \mbox{ 300 eV}$, $\Gamma=300$ with mildly
  relativistic electrons with initial $\gamma_{e} ^{\prime} \sim 2$,
  $\tau_{\mathrm{initial}} = 5$, and $N_{\mathrm{rh}} = $ 10, 100,
  1000 electron re-heating events. Above $E_{\mathrm{pk}}$, 
  $f_{\nu} \propto \nu^{-1.43}$ for $N_{\mathrm{rh}} = $ 100 for 
  $\sim 2$ decades in energy. \textit{Top-right Panel:} Same as
  top-left panel, but with $\gamma_{e} ^{\prime} \sim 30$ and
  $N_{\mathrm{rh}} = $ 2, 20, 200 electron re-heating events. 
  Above $E_{\mathrm{pk}}$, 
  $f_{\nu} \propto \nu^{-1.06}$ for $N_{\mathrm{rh}} = $ 20 for 
  $\sim 2$ decades in energy. \textit{Bottom-left Panel:} Simulation 
  results for the Comptonization of seed BB photons
  with $k_{B} T_{\gamma} ^{\prime} = \mbox{ 100 eV}$, $\Gamma=300$ with mildly
  relativistic electrons with initial $\gamma_{e} ^{\prime} \sim 2$,
  $\tau_{\mathrm{initial}} = 5$, and $N_{\mathrm{rh}} = $ 5, 50,
  500 electron re-heating events. Above $E_{\mathrm{pk}}$, 
  $f_{\nu} \propto \nu^{-1.41}$ for $N_{\mathrm{rh}} = $ 50 for 
  $\sim 2$ decades in energy. \textit{bottom-right Panel:} Same as
  top-left panel, but with $\gamma_{e} ^{\prime} \sim 50$ and
  $N_{\mathrm{rh}} = $ 0, 5, 50 electron re-heating events.
  Above $E_{\mathrm{pk}}$, 
  $f_{\nu} \propto \nu^{-1.00}$ for $N_{\mathrm{rh}} = $ 5 for 
  $\sim 2$ decades in energy. \label{fig5}}
\end{figure*}

In the top-left panel of Figure \ref{fig5}, we show electron
re-heating results for $\tau_{\mathrm{initial}} = 5$,
$k_{B} T_{\gamma} ^{\prime} = \mbox{ 300 eV}$,
mildly relativistic electrons with initial
$\gamma_{e} ^{\prime} \sim 2$, and
$N_{\mathrm{rh}} = $ 10, 100, 1000.
We considered a larger $\tau_{\mathrm{initial}}$ (5 as opposed to 2)
to allow a larger space for the re-heating events to occur.
For $N_{\mathrm{rh}} = 10$, $f_{\nu}$ still drops after the peak
energy by $\sim 2$ orders of magnitude. For $N_{\mathrm{rh}} =$ 100,
$f_{\nu}$ displays a power-law above the peak energy for
$\sim 2$ decades, with a spectrum $f_{\nu} \propto \nu^{-1.43}$.
For $N_{\mathrm{rh}} =$ 1000,
$f_{\nu}$ also displays a power-law above the peak energy for
$\sim 2$ decades, with a shallower spectrum $f_{\nu} \propto \nu^{-0.67}$.
In the top-right panel, we show simulations results for
$\tau_{\mathrm{initial}} = 5$, $k_{B} T_{\gamma} ^{\prime} = \mbox{ 300 eV}$,
$\gamma_{e} ^{\prime} \sim 30$, and $N_{\mathrm{rh}} = $ 2, 20, 200.
For $N_{\mathrm{rh}} = $ 20, 200, there is a power-law spectrum above
the peak energy for $\sim 2$ decades, with $f_{\nu} \propto \nu^{-1.06}$
for $N_{\mathrm{rh}} = $ 20 and $f_{\nu} \propto \nu^{-0.34}$
for $N_{\mathrm{rh}} = $ 200. In the bottom-left panel, we show
electron re-heating simulations for $\tau_{\mathrm{initial}} = 5$,
$k_{B} T_{\gamma} ^{\prime} = \mbox{ 100 eV}$,
$\gamma_{e} ^{\prime} \sim 2$. For $N_{\mathrm{rh}} = $ 50, 500,
there is a power-law spectrum above
the peak energy for $\sim 2$ decades, with $f_{\nu} \propto \nu^{-1.41}$
for $N_{\mathrm{rh}} = $ 50 and $f_{\nu} \propto \nu^{-0.59}$
for $N_{\mathrm{rh}} = $ 500. In the bottom-right panel, we show
electron re-heating simulations for $\tau_{\mathrm{initial}} = 5$,
$k_{B} T_{\gamma} ^{\prime} = \mbox{ 100 eV}$,
$\gamma_{e} ^{\prime} \sim 50$. For $N_{\mathrm{rh}} = $ 5, 50,
there is a power-law spectrum above
the peak energy for $\sim 2$ decades, with $f_{\nu} \propto \nu^{-1.00}$
for $N_{\mathrm{rh}} = $ 5 and $f_{\nu} \propto \nu^{-0.44}$
for $N_{\mathrm{rh}} = $ 50.

In summary, the main finding of considering electron re-heating is
that there is only a $f_{\nu} \propto \nu^{-1}$ spectrum above the
peak energy if a specific number if electron re-heating events are considered. If too few
electron re-heating events are considered, there is still a sharp drop
in $f_{\nu}$ after the peak energy and if too many electron re-heating
events are considered, the high-energy spectrum will be shallower than
$f_{\nu} \propto \nu^{-1}$. We also point out that the
re-heating simulations in Figure \ref{fig5} with a spectrum close to
$f_{\nu} \propto \nu^{-1}$ peak at $\sim \mbox{100 keV}$ and the
high-energy spectrum for these simulations extends to at least 10 MeV,
in agreement with the prompt emission observations.
%
%
%
%
\section{Discussion of Results for the Comptonization of
BB Photons}
\label{discussion_of_results}

In this section, we first discuss an energy requirement the electrons must
satisfy to have enough energy to transfer to the photons to produce a
power-law spectrum above the peak-energy. This energy requirement is a
necessary, but not a sufficient condition, to explain the production
of a power-law spectrum above the peak-energy. To give a more detailed
explanation for the simulation results, we calculate the number of photons upscattered to
higher energies in a simulation and compare this to the number of
photons needed to be upscattered to produce a
power-law spectrum. After this calculation, we discuss the interpretation for the drop
in $f_{\nu}$ by $\sim 2$ orders of magnitude, immediately after the
peak energy $E_{\mathrm{pk}}$, for the simulations with
$k_{B} T_{\gamma} ^{\prime} \sim \mbox{ 30 eV}-\mbox{ 300 eV}$,
one heating event, and $\tau_{\mathrm{initial}} = 2$.
We also discuss the interpretation of the
simulations with $k_{B} T_{\gamma} ^{\prime} = \mbox{ 30 eV}$
and $\tau_{\mathrm{initial}}$ 8, 16, which show a power-law spectrum
above $E_{\mathrm{pk}}$. We then apply this
interpretation to the electron re-heating simulations with
$k_{B} T_{\gamma} ^{\prime} = $ 100 eV, 300 eV to estimate how
many electron reheating events it takes to produce a power-law
spectrum above $E_{\mathrm{pk}}$. Lastly, we discuss the
dependence of the simulation results on the photon to electron ratio.

\subsection{Energy Requirement for Power-Law Spectrum}

In order to produce a power-law spectrum above the
peak-energy,  the electrons must
have enough energy to transfer to the photons so they can populate a
high-energy tail. A power-law spectrum
can develop if the energy of a significant
fraction of the photons near the BB peak-energy, $\sim 1/2$, can be
increased by a factor $\sim 2$. Taking the number of photons
near the BB peak to be $\sim N_{\gamma}$ (most of the photons are
near the peak), the energy requirement the electrons must satisfy is
\begin{equation}
N_{e} m_{e} c^{2} \gamma_{e} ^{\prime} \gtrsim 2 (N_{\gamma}/2) k_{B} T_{\gamma} ^{\prime}
\label{en_req_cond}
\end{equation}
When considering
$k_{B} T_{\gamma} ^{\prime} = $ 300 eV and
$k_{B} T_{\gamma} ^{\prime} = $ 100 eV, we considered
$\gamma_{e} ^{\prime} \sim 30$ and
$\gamma_{e} ^{\prime} \sim 50$, respectively. For these two cases,
for $N_{\gamma}/N_{e} \sim 10^{5}$ (as observed for the GRB prompt emission),
the electrons just meet the energy requirement, making the production
of a power-law spectrum difficult (in agreement with the results
presented in the top two panels of Figure\ref{fig4}). When considering
$k_{B} T_{\gamma} ^{\prime} = $ 30 eV, we considered
$\gamma_{e} ^{\prime} \sim 80$. For this case, the energy in the
electrons is $\sim 10$ larger than the energy that needs to be
transferred to the photons. Thus, the electrons have enough energy
to transfer to the photons to produce a power-law spectrum for this
case, in agreement with the results
presented in the bottom-left panel of Figure\ref{fig4}.

We now discuss a more detailed calculation to understand and
interpret out simulation results.

\subsection{Discussion of MC Simulation Results with One Heating
  Event}
\label{one_heating_event_discussion}

To understand our MC simulation results, we need to determine the number of
photons that need to be upscattered to energies larger than
$E_{\mathrm{pk}}$ to produce a power-law spectrum. We refer to this
quantity as $N_{\mathrm{pl}}$. We then compare
$N_{\mathrm{pl}}$ to the total number of photons that are upscattered to
energies larger than $E_{\mathrm{pk}}$ in a MC simulation, which is
given by the number of electrons in a simulation, $N_{e}$,
multiplied by $N_{\mathrm{Comp}}$. $N_{\mathrm{Comp}}$ represent the
number of scatterings it takes to cool an electron to a critical
$\gamma_{e} ^{\prime}$ at which Comptonization is no longer
important. Thus, in order to produce a power-law above
$E_{\mathrm{pk}}$, we need the condition
\begin{equation}
N_{e} N_{\mathrm{Comp}} \gtrsim N_{\mathrm{pl}}
\label{N_pl_condition}
\end{equation}
to be satisfied. We now estimate $N_{\mathrm{pl}}$ and
$N_{\mathrm{Comp}}$ for our MC simulations.

\subsubsection{Estimating $N_{\mathrm{pl}}$}
The number of photons in some bin with energy $E$, denoted as $N_{E}$,
is given by
\begin{equation}
N_{E} = \frac{ N_{\mathrm{pk}} }{ E_{\mathrm{pk}} }
  \left( \frac{E}{E_{\mathrm{pk}}} \right)^{-\beta-1} .
\end{equation}
In this equation, $N_{\mathrm{pk}}$ is the number of photons at
$E_{\mathrm{pk}}$ and the spectral index $\beta$ is defined in the
$f_{\nu}$ sense, i.e. $f_{\nu} \propto \nu^{-\beta}$.
To determine the total number of photons needed to produce a power-law
above $E_{\mathrm{pk}} $, we integrate $N_{E} dE$ (sum over
all the bins) from $E_{\mathrm{pk}} $ to infinity:
\begin{equation}
N_{\mathrm{pl}} = \int\limits_{E_{\mathrm{pk}} ^{\prime}}^{\infty} \frac{ N_{\mathrm{pk}} }{ E_{\mathrm{pk}} }
  \left( \frac{E}{E_{\mathrm{pk}} } \right)^{-\beta-1}  \, dE
  = \frac{N_{\mathrm{pk}}}{\beta} .
\end{equation}
For the prompt emission, the typical high-energy spectral index is
$\beta = 1.2$ \citep{preece_et_al_2000}.
Thus, we approximate $N_{\mathrm{pl}} \sim N_{\mathrm{pk}}$. Since the
majority of the photons in a simulation are near the peak of the BB
spectrum, we approximate
$N_{\mathrm{pl}} \sim N_{\mathrm{pk}} \sim N_{\gamma} $. With this result for
$N_{\mathrm{pl}}$, we can rewrite the condition to produce a power-law
above $E_{\mathrm{pk}} $ (Equation \ref{N_pl_condition}) in
terms of the photon to electron ratio ($N_{\gamma}/N_{e}$):
\begin{equation}
N_{\mathrm{Comp}} \gtrsim \frac{N_{\gamma}}{N_{e}}
\sim 1 \times 10^{5}.
\label{N_pl_condition_ph_el_ratio}
\end{equation}
In the above expression, we used
$N_{\gamma}/N_{e} = 10^{5}$
for our MC simulations (Equation \ref{photon_electron_ratio_eq}).

\subsubsection{Condition for electron $\gamma_{e} ^{\prime}$ at which Comptonization
is no longer important}
The Compton-$Y$ parameter determines if the energy of a photon will change
significantly after undergoing multiple scatterings with electrons in a
optically thick medium. The expression for the Compton-$Y$ parameter
is \citep{rybicki_and_lightman_1979}
\begin{equation}
Y = 2 \tau_{\mathrm{initial}} \times \mbox{max} \left[ \frac{4 k_{B}
    T_{e}^{\prime}}{m_{e} c^{2}}  , \frac{4}{3} [(\gamma_{e} ^{\prime}) ^{2} -1] \right] ,
\label{compton-y-eqn}
\end{equation}
where $2 \tau_{\mathrm{initial}}$ is the average number of scatterings
for each photon \citep{begue_et_al_2013} and the average fractional change of energy for a
photon after a scattering event is either $4 k_{B} T_{e} ^{\prime}/m_{e} c^{2}$
(for mildly-relativistic or sub-relativistic electrons)
or $(4/3)  [(\gamma_{e} ^{\prime}) ^{2} -1] $ (for relativistic electrons), depending on the
$\gamma_{e} ^{\prime}$ of the electron.
We take the transition from relativistic to mildly-relativistic
speeds to occur at $\gamma_{e} ^{\prime} = 2$.

The condition Compton-$Y \gtrsim 1$ needs to be satisfied for
Comptonization to be important.
At the start of our MC simulations,
Compton-$Y > 1$
since both $2 \tau_{\mathrm{initial}}$
(we consider $\tau_{\mathrm{initial}} \ge 2$) and
$\mbox{max} [ 4 k_{B} T_{e} ^{\prime}/m_{e} c^{2},
(4/3)  ([\gamma_{e} ^{\prime}] ^{2} -1) ]$ (we consider
$\gamma_{e} ^{\prime} \ge 2$) are larger than one.
However, as the simulations proceed, since $N_{\gamma} \gg N_{e}$,
the electrons may cool to a
point where Comptonization is no longer important.
The critical condition
at which Comptonization is no longer important is given by
\begin{equation}
2 \tau_{\mathrm{initial}} \times
\frac{4 k_{B} T_{e}^{\prime}}{m_{e} c^{2}} \sim 1 .
\end{equation}
Entering
$k_{B} T_{e} ^{\prime} \sim
m_{e} c^{2} (\gamma_{e, \mathrm{Comp}} ^{\prime} - 1) $,
where $\gamma_{e, \mathrm{Comp}} ^{\prime} $ is the electron Lorentz
factor at which Comptonization is no longer important, we find
\begin{equation}
\gamma_{e, \mathrm{Comp}} = 1 + \frac{1}{8 \tau_{\mathrm{initial}}} .
\label{gam_e_comp_eqn}
\end{equation}
Thus, $\gamma_{e, \mathrm{Comp}} = 1.06$ for $\tau_{\mathrm{initial}}=2$
and $\gamma_{e, \mathrm{Comp}} = 1.008$ for $\tau_{\mathrm{initial}}=16$.
With the value for $\gamma_{e, \mathrm{Comp}}$, we can now estimate
$N_{\mathrm{Comp}}$.

\subsubsection{Estimating $N_{\mathrm{Comp}}$}
The final $\gamma_{e} ^{\prime}$
($\gamma_{e, f} ^{\prime}$) of an electron after a scattering event can
be found from energy conservation
(see Appendix \ref{appen_elec_ener_dir_update}), which is given by
\begin{equation}
\gamma_{e, f} ^{\prime} =
\frac{E_{\gamma, i} ^{\prime} - E_{\gamma, f} ^{\prime} + m_{e} c^{2}
\gamma_{e, i} ^{\prime}}{m_{e} c^{2}} ,
\label{final_gam_e_IC_scat}
\end{equation}
where $E_{\gamma, i} ^{\prime}$ $[E_{\gamma, f} ^{\prime}]$
is the photon energy before [after] the scattering event
and $\gamma_{e, i} ^{\prime}$ is the
electron Lorentz factor before the scattering event.
As we discussed above, the average change of
energy for a photon after a scattering event depends on
whether $\gamma_{e} ^{\prime} > 2$ or
$\gamma_{e} ^{\prime} < 2$ \citep{rybicki_and_lightman_1979}:
\begin{alignat}{3}
E_{\gamma, f} ^{\prime} - E_{\gamma, i} ^{\prime}  &=
\frac{4}{3} [ (\gamma_{e, i} ^{\prime})^{2} - 1] E_{\gamma, i}^{\prime}  &
\mbox{,     for } \gamma_{e,i} ^{\prime} > 2 \\
E_{\gamma, f} ^{\prime} - E_{\gamma, i} ^{\prime}  &=
\frac{4 k_{B} T_{e} ^{\prime}}{m_{e} c^{2}} E_{\gamma, i} ^{\prime} &
\mbox{,     for } \gamma_{e,i} ^{\prime} < 2 .
\end{alignat}
Substituting these expressions for
$E_{\gamma, f} ^{\prime} - E_{\gamma, i} ^{\prime} $ into
Equation \ref{final_gam_e_IC_scat}, we can solve for the change in
electron $\gamma_{e} ^{\prime}$ after a scattering event in terms of
$E_{\gamma, i} ^{\prime}$ and $\gamma_{e, i} ^{\prime}$:
\begin{alignat}{3}
\gamma_{e, f} ^{\prime} - \gamma_{e, i} ^{\prime} &=
- \frac{4 E_{\gamma, i} ^{\prime}}{3m_{e} c^{2}}
[ (\gamma_{e, i} ^{\prime})^{2} - 1] &
\mbox{,     for } \gamma_{e,i} ^{\prime} > 2
\label{gam_e_f_R} \\
\gamma_{e, f} ^{\prime} - \gamma_{e, i} ^{\prime} &=
- \frac{4 E_{\gamma, i} ^{\prime}}{m_{e} c^{2}} [ \gamma_{e, i} ^{\prime} - 1] &
\mbox{,     for } \gamma_{e,i} ^{\prime} < 2 .
\label{gam_e_f_MR}
\end{alignat}
In Equation \ref{gam_e_f_MR}, for $\gamma_{e,i} ^{\prime} < 2$,
we used $k_{B} T_{e} ^{\prime} \sim m_{e} c^{2} (\gamma_{e, i} ^{\prime} - 1)$
for the electron temperature.
If we define the change in electron
$\gamma_{e} ^{\prime}$ per scattering as
$d \gamma_{e} ^{\prime}/d N$,
we can re-write
Equation \ref{gam_e_f_R} and
Equation \ref{gam_e_f_MR} in differential form:
\begin{alignat}{3}
\frac{d \gamma_{e} ^{\prime}}{dN}  &=
- \frac{4 E_{\gamma} ^{\prime}}{3m_{e} c^{2}}
[ (\gamma_{e} ^{\prime})^{2} - 1] &
\mbox{,     for } \gamma_{e} ^{\prime} > 2 \\
\frac{d \gamma_{e} ^{\prime}}{dN}  &=
- \frac{4 E_{\gamma} ^{\prime}}{m_{e} c^{2}} [ \gamma_{e} ^{\prime} - 1] &
\mbox{,     for } \gamma_{e} ^{\prime} < 2 .
\end{alignat}
In the above expressions, $E_{\gamma} ^{\prime}$ represents the energy
of the photon in the jet-comoving frame before the scattering event.
Solving these differential equations, we find
\begin{alignat}{3}
N_{\mathrm{Cool, Rel}} &=
\frac{3 m_{e} c^{2}}{8 E_{\gamma} ^{\prime}}
\mbox{ ln} \left[ \frac{(\gamma_{e,\mathrm{MR}} ^{\prime} + 1)
(\gamma_{e, i} ^{\prime} -1)}{(\gamma_{e,\mathrm{MR}} ^{\prime} - 1)
(\gamma_{e, i} ^{\prime} +1)} \right] &
\mbox{,     for } \gamma_{e,i} ^{\prime} > 2
\label{n_cool_rel_analytic} \\
N_{\mathrm{Cool, MR}}  &=
\frac{m_{e} c^{2}}{4 E_{\gamma} ^{\prime}}
\mbox{ ln} \left[ \frac{(\gamma_{e, i} ^{\prime} -1)}{(\gamma_{e,\mathrm{Comp}} ^{\prime} - 1)} \right] &
\mbox{,     for } \gamma_{e,i} ^{\prime} \le 2 .
\label{n_cool_MR_analytic}
\end{alignat}
In Equation \ref{n_cool_rel_analytic},
$N_{\mathrm{Cool, Rel}}$ represents the number of scatterings it takes
to cool a relativistic electron with $\gamma_{e,i} ^{\prime} > 2$
to $\gamma_{e,\mathrm{MR}} ^{\prime} = 2$.
After an electron cools
below $\gamma_{e,\mathrm{MR}} ^{\prime} = 2$,
$N_{\mathrm{Cool, MR}}$ represents the number of scatterings it takes
to cool a mildly-relativistic (MR) electron from
$\gamma_{e,i} ^{\prime} \le 2$ to $\gamma_{e,\mathrm{Comp}} ^{\prime}$.

For our simulations, we considered
$\gamma_{e,i} ^{\prime} \ge 2$. If we consider
$\gamma_{e,i} ^{\prime} > 2$, to compute $N_{\mathrm{Comp}}$, we first need to compute the number of
scatterings it takes to cool an electron to
$\gamma_{e,\mathrm{MR}} ^{\prime} = 2$
(Equation \ref{n_cool_rel_analytic}). Then, we need to compute the
number of scattering it takes to cool an electron with
$\gamma_{e,i} ^{\prime} = 2$ to $\gamma_{e,\mathrm{Comp}} ^{\prime}$
(Equation \ref{n_cool_MR_analytic}).
Thus, the number of scatterings it takes to cool a relativistic
electron to $\gamma_{e,\mathrm{Comp}} ^{\prime}$
(defined as $N_{\mathrm{Comp, Rel}}$) is
\begin{equation}
N_{\mathrm{Comp, Rel}} =
N_{\mathrm{Cool, Rel}} (\gamma_{e, i} ^{\prime} > 2) +
N_{\mathrm{Cool, MR}} (\gamma_{e, i} ^{\prime}= 2) .
\label{n_Comp_rel}
\end{equation}
On the other hand, for a mildly-relativistic electron with $\gamma_{e,i} ^{\prime} \le 2$,
the number of scatterings it takes to cool an electron
from $\gamma_{e,i} ^{\prime}$ to
$\gamma_{e,\mathrm{Comp}} ^{\prime}$
(defined as $N_{\mathrm{Comp, MR}}$) is simply found by computing
$N_{\mathrm{Cool, MR}}$ (Equation \ref{n_cool_MR_analytic}).
We give a summary of the values of $N_{\mathrm{Comp}}$ for the
simulations we presented in Figures \ref{fig2}-\ref{fig4} in
Table \ref{table2}.
\begin{table*}
\centering
\begin{tabular}{ c c c c }
     \hline
     & $N_{\mathrm{Cool, MR}} (\gamma_{e,i} ^{\prime} = 2)$ &
     $N_{\mathrm{Cool, MR}} (\gamma_{e,i} ^{\prime} = 2)$ &
     $N_{\mathrm{Cool, Rel}} (\gamma_{e, i} ^{\prime} > 2)$ \\
     & $\tau_{\mathrm{initial}} = 2 $ &
     $\tau_{\mathrm{initial}} = 16 $ & \\ \hline
     $k_{B} T_{\gamma} ^{\prime} = $30 eV & $\sim 10^{4}$ &
     $\sim 2 \times 10^{4}$ &
     $\sim 7 \times 10^{3}$ for $\gamma_{e, i}=80$ \\ \hline
    $k_{B} T_{\gamma} ^{\prime} = $100 eV & $\sim 3 \times 10^{3}$ &
    $\sim 6 \times 10^{3}$ &
    $\sim 2 \times 10^{3}$ for $\gamma_{e, i}=50$ \\ \hline
    $k_{B} T_{\gamma} ^{\prime} = $300 eV & $\sim 10^{3}$ &
    $\sim 2 \times 10^{3}$ &
     $\sim 7 \times 10^{2}$ for $\gamma_{e, i}=30$ \\ \hline
\end{tabular}
\caption{Values of $N_{\mathrm{Comp}}$ for the simulations we
  presented in Figures \ref{fig2}-\ref{fig4}.
  $N_{\mathrm{Comp, MR}} $ and $N_{\mathrm{Comp, Rel}}$ were
  calculated with Equation \ref{n_cool_MR_analytic} and
  Equation \ref{n_Comp_rel}, respectively. \label{table2}}
\end{table*}
When using
Equation \ref{n_cool_rel_analytic} and \ref{n_cool_MR_analytic}
to calculate
$N_{\mathrm{Cool, Rel}}$ and $N_{\mathrm{Cool, MR}}$,
we assume $E_{\gamma} ^{\prime} \sim k_{B} T_{\gamma} ^{\prime}$,
i.e. that the electrons cool mostly by scattering photons near the BB
peak. This is not a bad approximation since most of the photons are
near the BB peak.

\subsubsection{Interpretation of MC Simulation Results with One Heating
  Event}
\label{one_heat_event_interp}

From the estimates we provided in Table \ref{table2}
for $N_{\mathrm{Comp}}$ at $\tau_{\mathrm{initial}} = 2$,
for $k_{B} T_{\gamma} ^{\prime} \sim  30 \mbox{ eV} - 300 \mbox{ eV}$
and initial $\gamma_{e} ^{\prime} \sim 2-80$,
$N_{\mathrm{Comp}} \sim 2000-10000 \ll (N_{\gamma}/N_{e})$
(Equation \ref{N_pl_condition_ph_el_ratio}).
Thus, not enough photons are upscattered to energies
above $E_{\mathrm{pk}} $ to produce a power-law
spectrum. The fraction of photons that can be upscattered to energies
larger than $E_{\mathrm{pk}}$ is given by
\begin{equation}
\mbox{Fraction of upscattered photons } =
\frac{N_{e} N_{\mathrm{Comp}}}{ N_{\gamma}} =
\frac{ N_{\mathrm{Comp}}}{(N_{\gamma}/N_{e})} ,
\label{fract_of_upscatt_photons}
\end{equation}
where $N_{e} N_{\mathrm{Comp}}$ is the total number of photons
upscattered to energies larger that $E_{\mathrm{pk}}$
and $N_{\gamma}$ is the number of photons
near the peak of the BB spectrum.
For $\tau_{\mathrm{initial}} = 2$,
$ N_{\mathrm{Comp}}/[N_{\gamma}/N_{e}] \sim 10^{-2}$.
Since only $\sim 10^{-2}$ of the photons near the BB peak
are upscattered to higher energies, this result explains why $f_{\nu}$
drops by $\sim 2$ orders of
magnitude and then a power-law spectrum begins to
develop (Figures \ref{fig2}-\ref{fig3}).

\begin{figure}
\centering
\includegraphics[scale=0.45]{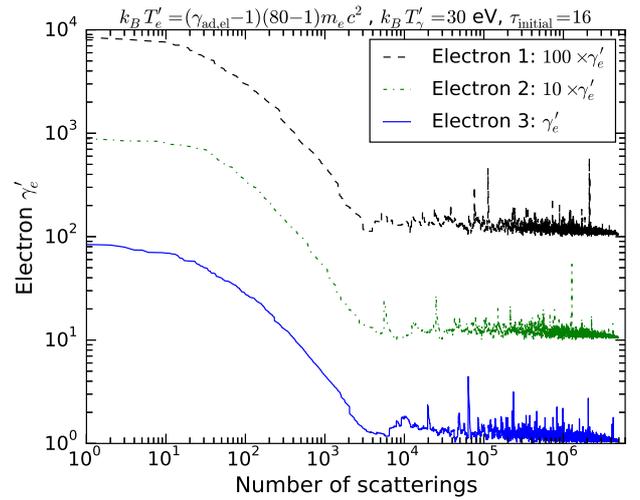}
\caption{Evolution of $\gamma_{e} ^{\prime}$ for 3 electrons from the
  $\tau_{\mathrm{initial}} = 16$ simulations shown in the Bottom-Left
  panel of Figure \ref{fig4}. The $\gamma_{e} ^{\prime}$ of the electrons were
  offset by a factor of 10 to better see the evolution of
  $\gamma_{e} ^{\prime}$ for each electron. Each of the spikes for
  $\gamma_{e} ^{\prime}$ represents an episode when an electron
  interacts with an energetic photon, causing the energy of the
  electron to increase by a large factor.
  \label{fig6} }
\end{figure}

We now apply our analytical estimates to the
simulations results for $\tau_{\mathrm{initial}} = 16$
presented in Figure \ref{fig4}.
For $k_{B} T_{\gamma} ^{\prime} \sim  100 \mbox{ eV} - 300 \mbox{ eV}$
and $\tau_{\mathrm{initial}} = 16$,
$N_{\mathrm{Comp}} \sim 3000-6000 < N_{\gamma}/N_{e}$ (Table \ref{table2}).
Thus, we do not expect a power-law
spectrum to form above the peak energy, in agreement with the simulation results in
Top-Left and Top-Right panels of Figure \ref{fig4}.
For $k_{B} T_{\gamma} ^{\prime} = \mbox{ 30 eV}$ and
$\tau_{\mathrm{initial}} = 16$,
$N_{\mathrm{Comp}} \sim 2 \times 10^{4}$ (Table \ref{table2}).
Since $N_{\mathrm{Comp}} $ is less that $N_{\gamma}/N_{e}$ by a factor of
5, we do not expect a power-law spectrum to form
above $E_{\mathrm{pk}}$. However, our analytical estimate assumes that
the electrons only cool, and does not consider the possibility that an
electron can gain energy by interacting with an energetic photon.
If the $\gamma_{e} ^{\prime}$ of the electron
rises to $\gamma_{e} ^{\prime} \sim 2$, it can upscatter $\sim 20000$
more photons to higher energies since
$N_{\mathrm{Cool, MR}} \sim 20000$ for
$k_{B} T_{\gamma} ^{\prime} = \mbox{  30 eV}$,
$\gamma_{e} ^{\prime} = 80$, and
$\tau_{\mathrm{initial}} = 16$ (Table \ref{table2}).
In Figure \ref{fig6}, we show the evolution of
$\gamma_{e} ^{\prime}$ for 3 electrons in the simulation.
Initially, the 3 electrons cool from $\gamma_{e} ^{\prime} = 80$ to
$\gamma_{e} ^{\prime} \sim 1$. However, there are $\sim 5$ instances where the
$\gamma_{e} ^{\prime}$ of the electron rises to
$\gamma_{e} ^{\prime} \sim 2$. Thus, the number of photons that are upscattered
to larger energies is increased by $\sim 5$, which give us
$N_{\mathrm{Comp}} \sim N_{\gamma}/N_{e}$
(Equation \ref{N_pl_condition_ph_el_ratio}), and explains why the
$\tau_{\mathrm{initial}} = 16$
simulations in the Bottom-Left panel of Figure \ref{fig4}
show a power-law spectrum above $E_{\mathrm{pk}}$.

\subsection{Discussion of MC Simulation Results with Electron Reheating}

In this subsection, we use the results for $N_{\mathrm{pl}}$ and
$N_{\mathrm{Comp}}$ to understand the
electron re-heating simulations with
$k_{B} T_{\gamma} ^{\prime} = $ 100 eV - 300 eV presented in
Figure \ref{fig5}. We estimate
$N_{\mathrm{rh, min}}$, the minimum number of re-heating events
needed to produce a power-law spectrum above
$E_{\mathrm{pk}}$.

\subsubsection{Estimating $N_{\mathrm{rh, min}}$}
With $N_{\mathrm{rh}}$
electron re-heating events, the number of photons that can be
upscattered to higher energies is
$\sim N_{\mathrm{rh}} N_{e} N_{\mathrm{Comp}}$.
$N_{\mathrm{rh, min}}$ is found by the condition where just enough
photons are upscattered to energies larger than
$E_{\mathrm{pk}} $ to produce a power-law spectrum, i.e.
\begin{equation}
N_{\mathrm{rh, min}} N_{e} N_{\mathrm{Comp}} = N_{\mathrm{pl}} .
\end{equation}
Using the result $N_{\mathrm{pl}} \sim N_{\gamma}$, the condition for
$N_{\mathrm{rh, min}}$ can be re-written in terms of the
photon to electron ratio:
\begin{equation}
N_{\mathrm{rh, min}} \sim
\frac{N_{\gamma}/N_{e}}{N_{\mathrm{Comp}}} .
\end{equation}
For $k_{B} T_{\gamma} ^{\prime} = \mbox{ 300 eV}$ and
$\gamma_{e, i} ^{\prime} \sim 2-30$,
$N_{\mathrm{Comp}} \sim 2000-3000$ (Table \ref{table2}); thus we
estimate $N_{\mathrm{rh, min}} \sim 30-50$. $N_{\mathrm{rh, min}}$ is within a factor
of a few of the simulations in the top panels of Figure \ref{fig5}, which show
a power-law above the peak-energy for $N_{\mathrm{rh}} \sim 20-100$.
For $k_{B} T_{\gamma} ^{\prime} = \mbox{ 100 eV}$ and
$\gamma_{e, i} ^{\prime} \sim 2-50$,
$N_{\mathrm{Comp}} \sim 5000-8000$ (Table \ref{table2}); thus we
estimate $N_{\mathrm{rh, min}} \sim 10-20$. $N_{\mathrm{rh, min}}$ is within a factor
of a few of the simulations in the bottom panels of Figure \ref{fig5}, which show
a power-law above the peak-energy for $N_{\mathrm{rh}} \sim 5-50$.

\subsection{Dependence of Comptonization of Seed BB Simulation Results on $N_{\gamma}/N_{e}$}

\begin{figure*}
\begin{center}$
\begin{array}{c}
\begin{array}{cc}
\hspace{-10mm}
\includegraphics[scale=0.45]{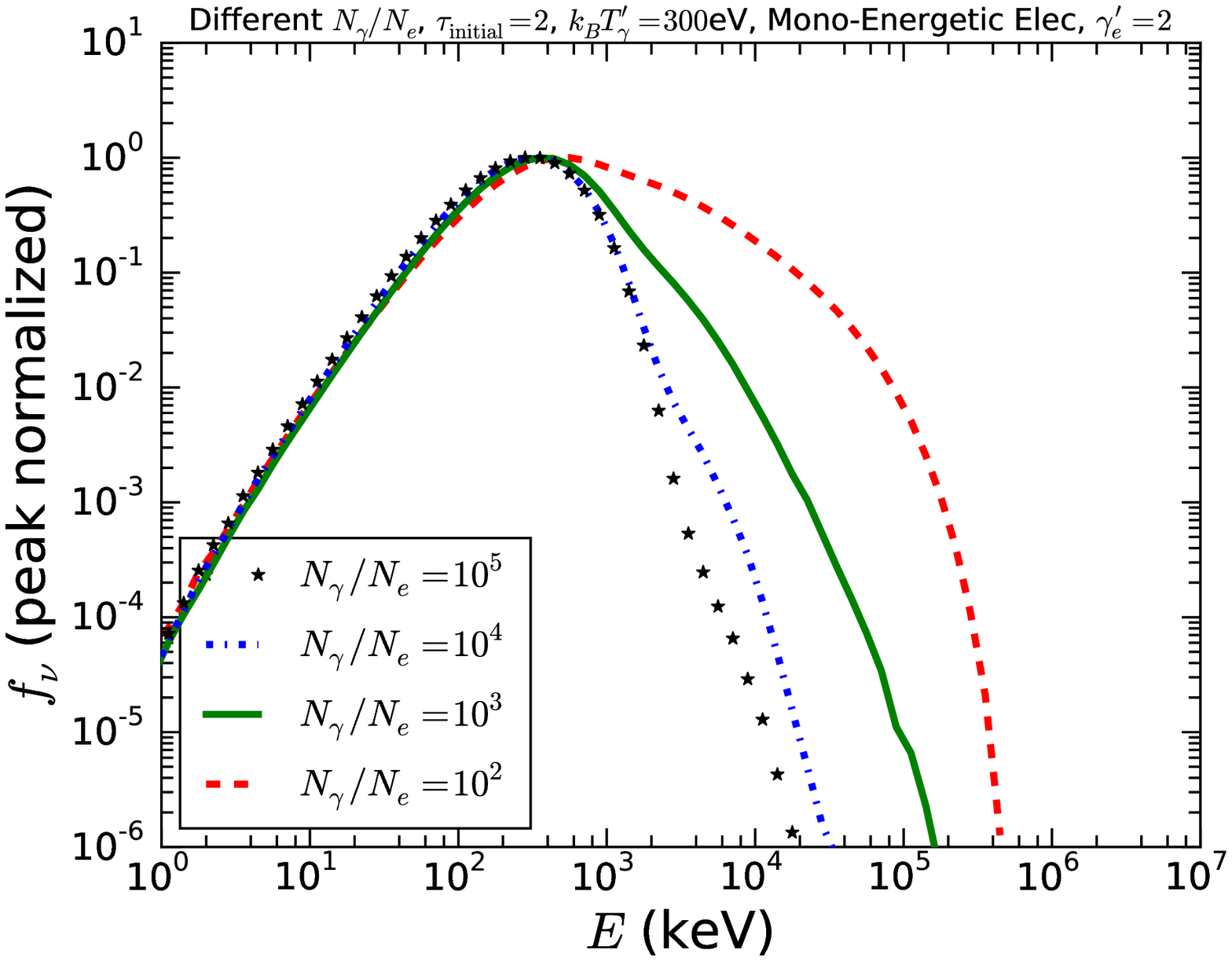} &
\hspace{0mm}
\includegraphics[scale=0.45]{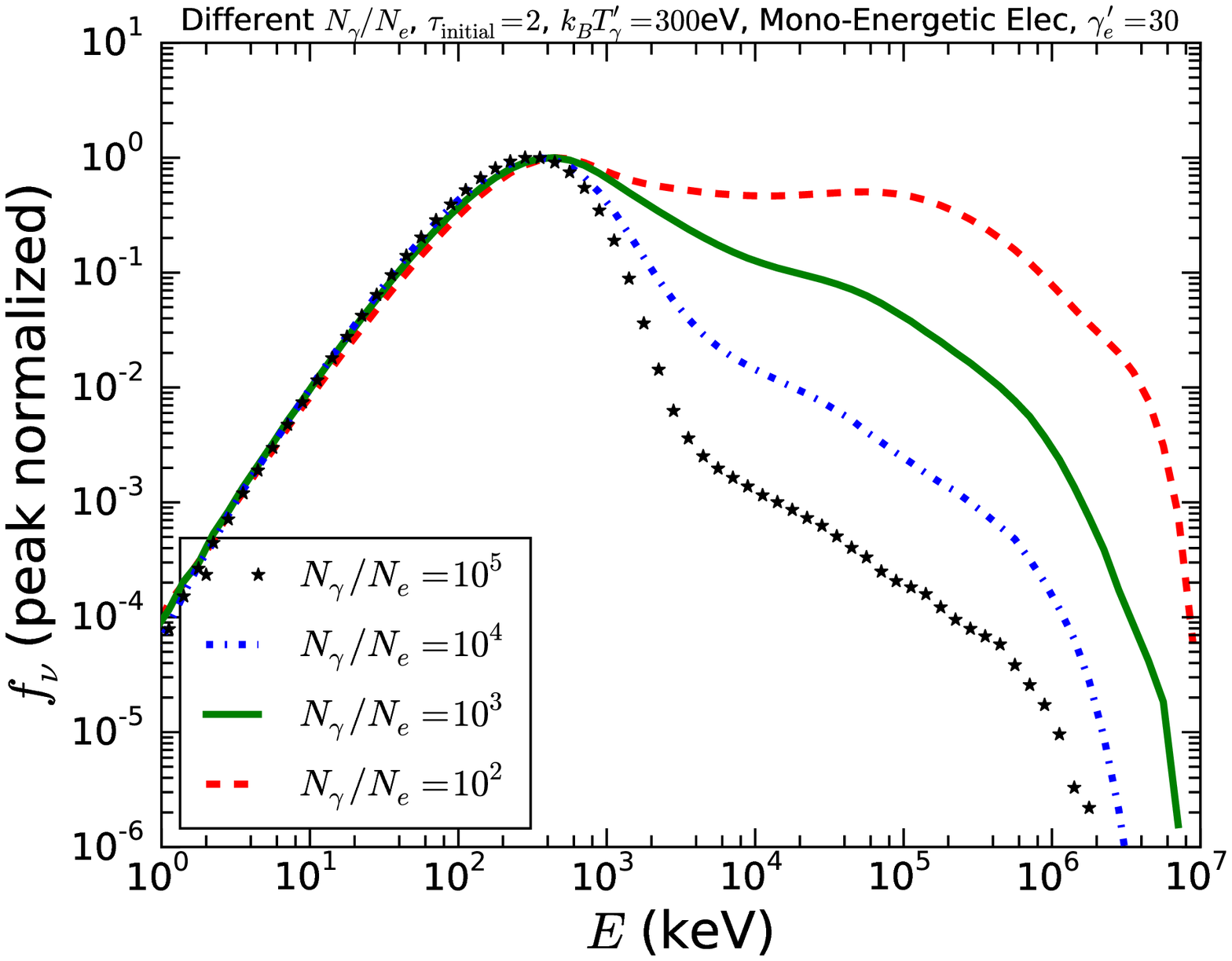}
\end{array}
\end{array}$
\end{center}
\caption{\textit{Left Panel:} Simulation results for the Comptonization of seed BB photons
  with $k_{B} T_{\gamma} ^{\prime} = \mbox{ 300 eV}$, $\Gamma =300$ with
  mildly relativistic electrons with initial $\gamma_{e} ^{\prime} = 2$,
  four different values for the photon to electron ratio,
  and $\tau_{\mathrm{initial}} = 2$. \textit{Right Panel:}
  Same as Left Panel but with mono-energetic electrons with
  initial $\gamma_{e} ^{\prime} = 30$. \label{fig7}}
\end{figure*}

In this subsection, we perform a quick set of simulations at
$\tau_{\mathrm{initial}} = 2$ to explore the dependence of the
simulation results on $N_{\gamma} / N_{e}$.
In the left panel (right panel) of Figure \ref{fig7}, we show simulations results for
$k_{B} T_{\gamma} ^{\prime} = \mbox{ 300 eV}$, $\Gamma=300$,
$N_{\gamma} / N_{e} = 10^{2}, 10^{3}, 10^{4}, 10^{5}$,
$\tau_{\mathrm{initial}} = 2$, and mono-energetic electrons with
initial $\gamma_{e} ^{\prime} = 2$ ($\gamma_{e} ^{\prime} = 30$).
For $N_{\gamma} / N_{e} = 10^{2}$ we considered $N_{e}=10^{6}$, for
$N_{\gamma} / N_{e} = 10^{3}$ we considered $N_{e}=10^{5}$, etc. This
was done to keep $N_{\gamma}=10^{8}$ so that the simulation output spectrum
can have more photons and thus a higher signal to noise.
For initial $\gamma_{e} ^{\prime} = 2$ (left panel of Figure \ref{fig7}), the
spectrum shows a sharp drop above the peak energy for
$N_{\gamma}/N_{e} = 10^{4}$. For $N_{\gamma}/N_{e} = 10^{3}$,
the spectrum shows a power-law with $f_{\nu} \propto \nu^{-1.60}$ and for
$N_{\gamma}/N_{e} = 10^{2}$ the spectrum also shows a power-law with
$f_{\nu} \propto \nu^{-0.80}$. For initial $\gamma_{e} ^{\prime} = 30$ (right panel of Figure \ref{fig7}), the
spectrum shows a sharp drop above the peak energy for
$N_{\gamma}/N_{e} = 10^{4}$. For $N_{\gamma}/N_{e} = 10^{3}$,
there is a power-law spectrum
$f_{\nu} \propto \nu^{-0.70}$ and for $N_{\gamma}/N_{e} = 10^{2}$
the spectrum is very shallow.
The difference in the simulation results
with $N_{\gamma}/N_{e}$ can be understood from a energetics
perspective. Equation \ref{en_req_cond}, the minimum energy the
electrons must have to transfer to the photons to produce a power-law
spectrum, can be rewritten in terms of the photon to electron ratio:
\begin{equation}
\frac{m_{e} c^{2} \gamma_{e} ^{\prime}}{N_{\gamma}/N_{e}} \gtrsim k_{B} T_{\gamma} ^{\prime} .
\end{equation}
Thus, for a fixed $k_{B} T_{\gamma} ^{\prime}$,
an increase in $N_{\gamma}/N_{e}$ makes it more difficult to produce a
power-law spectrum above the peak-energy, as demonstrated in both
panels of Figure \ref{fig7}. These results
highlight the strong dependence of the simulation results on
$N_{\gamma}/N_{e}$ and the importance of performing simulations with
$N_{\gamma}/N_{e} = 10^{5}$.

Lastly, we note that
\citet{lazzati_and_begelman_2010} also show a couple of simulation
results for mono-energetic electrons with $\gamma_{e} ^{\prime} = 2$, $\tau_{\mathrm{initial}} =
2$, and $N_{\gamma}/N_{e} = 10^{4}$ in their Figures 5-6. Their results for
$N_{\gamma}/N_{e} = 10^{4}$ also show a significant dip above the peak
energy by a factor $\sim 50$. However, our simulation results
in the left panel of Figure \ref{fig7} for $\gamma_{e} ^{\prime} = 2$ with
$N_{\gamma}/N_{e} = 10^{4}$ display a drop in $f_{\nu}$ above
$E_{\mathrm{pk}}$ by a factor $\sim 100$. This difference in the dip in $f_{\nu}$
above the peak energy can be explained by the fact
that \citet{lazzati_and_begelman_2010} considered a value
for $k_{B} T_{\gamma} ^{\prime}$ smaller than the $k_{B} T_{\gamma} ^{\prime}$ value we considered
by a factor of 2. Since
$N_{\mathrm{Cool, MR}} \propto 1/E_{\gamma} ^{\prime}$
(Equation \ref{n_cool_MR_analytic}),
a smaller value for $k_{B} T_{\gamma} ^{\prime}$ by a factor of 2
implies that twice as many photons will be upscattered to larger
energies (Equation \ref{fract_of_upscatt_photons}).
Thus, the $N_{\gamma}/N_{e} = 10^{4}$ presented in
\citet{lazzati_and_begelman_2010} are consistent with our analytical
estimates and with our simulation results for
$N_{\gamma}/N_{e} = 10^{4}$.

%
%
%

\section{Comptonization of Synchrotron $ f_{\nu}
  \propto \nu^{-1/2}$ Spectrum}
\label{sync_comp_sect}

In the previous sections, we considered the scenario where 
all the electrons were initially
piled-up at $\gamma_{e} ^{\prime} = \gamma_{\mathrm{SA}}$,
where the optical depth to synchrotron self-absorption is 1. However,
even when a system is optically thick to Thomson scattering as we
consider, it is possible for the system to be optically-thin to
synchrotron absorption. In this section, we consider a scenario where
the synchrotron emission is initially optically thin. Then, we use our
code to study how Comptonization modifies the photon spectrum produced
by the synchrotron emission.

The situation we consider is the
following. Initially, we take the electrons to be accelerated to a
power-law distribution $(\gamma_{e} ^{\prime}) ^{-p}$ with 
$\gamma_{i} ^{\prime} \gg \gamma_{\mathrm{SA}} ^{\prime}$, where $\gamma_{i} ^{\prime}$ is the
electron Lorentz factor where the power-law begins. Unless the
magnetic field is very small, for typical GRB parameters, the
electrons are expected to be in the fast cooling regime. In this
regime, the electrons produce a
$f_{\nu} \propto \nu^{-1/2}$ synchrotron spectrum \citep{ghisellini_et_al_2000}. 
The electrons then continue to cool by emitting synchrotron 
photons until their Lorentz factor
becomes $\gamma_{\mathrm{SA}} ^{\prime}$. For 
electron Lorentz factors 
$\gamma_{e} ^{\prime} \le \gamma_{\mathrm{SA}} ^{\prime}$,
although the electrons cannot cool via synchrotron emission, they can
cool by IC scattering photons. To model this scenario, we consider a seed photon spectrum 
$f_{\nu} \propto \nu^{-1/2}$, instead of a BB spectrum. We then
perform MC simulations to study how Comptonization
modifies a $f_{\nu} \propto \nu^{-1/2}$ photon spectrum. As before,
we take all the electrons to initially be piled-up at 
$\gamma_{e} ^{\prime} = \gamma_{\mathrm{SA}} ^{\prime}$.
In the next subsection, we describe the input
parameters we consider for these simulations.

\subsection{Input Parameters for Simulations with Seed $f_{\nu}
  \propto \nu^{-1/2}$ Spectrum}

For most of the input parameters, we considered the same values as
those we considered for the Comptonization of a seed BB
spectrum (discussed in Section \ref{input_param_sect}).
We considered $\Gamma = 300$,
$L = 10^{52} \mbox{ ergs}/\mbox{sec}$,
 $\tau_{\mathrm{initial}} = $ 2, 5, 8, 16,
$N_{e} = 10^{3}$,
$N_{\gamma} = 10^{8}$ to reach
$N_{\gamma}/N_{e} = 10^{5}$, and
$N_{\mathrm{collect}} =  N_{\gamma}/3$.
For the $f_{\nu} \propto \nu^{-1/2}$ seed spectrum, the input
parameters are $E_{1, \gamma} ^{\prime}$ and
$E_{2, \gamma} ^{\prime}$, the energy where the
$f_{\nu} \propto \nu^{-1/2}$ spectrum begins and ends, respectively,
in the jet-comoving frame. Thus, in the observer frame,
the $f_{\nu} \propto \nu^{-1/2}$ spectrum begins and ends at
$E_{1, \gamma} = \mathcal{D} E_{1, \gamma} ^{\prime}$ and
$E_{2, \gamma} = \mathcal{D} E_{2, \gamma} ^{\prime}$. As we will discuss
in the next subsection, the effect of Comptonizing a
$f_{\nu} \propto \nu^{-1/2}$ spectrum is to flatten it to
$f_{\nu} \propto \nu^{0}$, reminiscent of the low-energy spectrum of
the prompt emission. Since the observed $f_{\nu} \propto \nu^{0}$
spectrum for the prompt emission extends from $\lesssim \mbox{ 10 keV}$ to
$E_{\mathrm{pk}} = \mbox{ 300 keV}$, we considered 
$E_{1, \gamma} ^{\prime} = \mbox{ 0.01 eV}$
and $E_{2, \gamma} ^{\prime} = \mbox{ 300 eV}$. For the
electron distribution, we considered MB
electrons with
$k_{B} T_{e} ^{\prime} \sim (2000-1) m_{e} c^{2}$. 
We obtained $\gamma_{e} ^{\prime} \sim 2000$ by setting 
$\tau_{\mathrm{sync}} ^{\mathrm{MB}} = 1$ in Equation
\ref{opt_depth_sync_SA}. For $E_{\gamma} ^{\prime}$ 
in Equation \ref{opt_depth_sync_SA}, we took 0.01 eV since
most of the photons are at this energy for 
the $f_{\nu} \propto \nu^{-1/2}$ photon spectrum we consider.
Lastly, we note that we do not consider electron re-heating, i.e. the
electrons are only accelerated once at the start of the simulations.

\subsection{Simulation Results for Comptonization of $f_{\nu}
  \propto \nu^{-1/2}$ Seed Spectrum}

\begin{figure}
\centering
\includegraphics[scale=0.45]{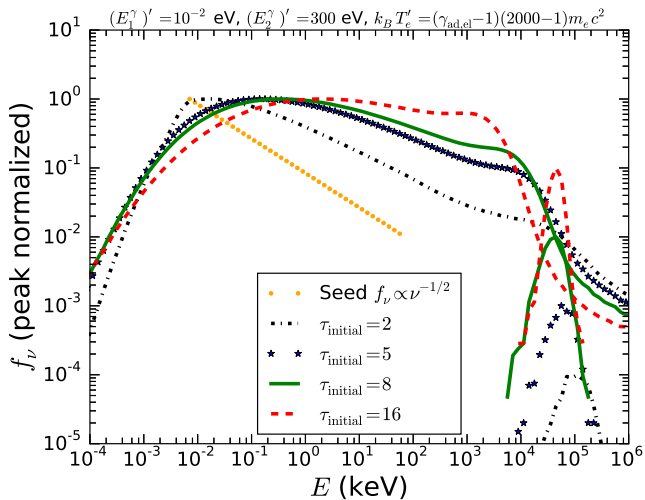}
\caption{Simulation results for the Comptonization of a seed
  $f_{\nu} \propto \nu^{-1/2}$ spectrum
  ($E_{1, \gamma} ^{\prime} = 10^{-2} \mbox{ eV}$, $E_{2, \gamma} ^{\prime} = \mbox{ 300 eV}$)
  with MB electrons with $k_{B} T_{e} ^{\prime} \sim (2000-1) m_{e} c^{2}$
  and $\tau_{\mathrm{initial}} = 2$, 5, 8, 16. As in Figure
  \ref{fig4}, the photon spectra are peak normalized and the electron spectra 
  (spectra in the bottom right) are shifted by a factor of
  10 for each $\tau_{\mathrm{initial}}$. \label{fig8} }
\end{figure}

In Figure \ref{fig8}, we show the results for
the Comptonization of a seed $f_{\nu} \propto \nu^{-1/2}$ spectrum
with $E_{1, \gamma} ^{\prime} = 0.01 \mbox{ eV}$,
$E_{2, \gamma} ^{\prime} = 300 \mbox{ eV}$, MB electrons with
$k_{B} T_{e} ^{\prime} \sim (2000-1) m_{e} c^{2}$, and
$\tau_{\mathrm{initial}} = $ 2, 5, 8, 16.
In addition, we also plot
the energy spectrum  $f_{\nu} = E N_{E}$ of the kinetic energies of
the electrons in the observer frame,
i.e. $m_{e} c^{2} (\gamma_{\mathrm{ad, el}} - 1) (\gamma_{e} ^{\prime} -1) \Gamma$, at the end of each
simulation. We will first discuss the results for the electron
distributions. At the end of the
$\tau_{\mathrm{initial}} = 2$ simulations, the electron distributions
peak at $k_{B} T_{e} ^{\prime} \sim 100 \mbox{ keV}$ (in the
jet-comoving frame). As $\tau_{\mathrm{initial}}$ increases,
$k_{B} T_{e} ^{\prime}$ doesn't change by much;
at the end of the $\tau_{\mathrm{initial}} = 16$ simulation
$k_{B} T_{e} ^{\prime} \sim 50 \mbox{ keV}$.
On the other hand, the photon spectrum
changes significantly as $\tau_{\mathrm{initial}}$ increases.
At the end of the $\tau_{\mathrm{initial}} = 2$ simulations, most of the
photons are still near $E_{1, \gamma}$, where most of the photons in
the seed spectrum are initially present. However, as
$\tau_{\mathrm{initial}}$ increases, more and more photons begin to be
upscattered to energies $\sim 10^{4} \mbox{ keV}$.
At the end of the $\tau_{\mathrm{initial}} = 16$,
the spectrum is flat from $\sim \mbox{1 keV}$ to
$\sim 10^{4} \mbox{ keV}$. After $\sim 10^{4} \mbox{ keV}$,
the spectrum declines rapidly for all of the simulations. After the
rapid decline, there are a significant number of photons with energies
$\gtrsim 10^{5}$ keV for all of the simulations. These photons remain
at these energies since they are highly Klein-Nishina suppressed and
cannot transfer their energy back to the electrons.
We now discuss the basic interpretation for the flattening of the
spectrum as $\tau_{\mathrm{initial}}$ increases. From
Equation \ref{gam_e_comp_eqn}, for $\tau_{\mathrm{initial}} = 16$,
the $\gamma_{e} ^{\prime}$ at which Comptonization is no longer important is
$\gamma_{e, \mathrm{Comp}} = 1.008$. At the end of the
$\tau_{\mathrm{initial}} = 16$ simulations,
$k_{B} T_{e} ^{\prime} \sim \mbox{ 50 keV}$. Using
$k_{B} T_{e} ^{\prime} \sim m_{e} c^{2} (\gamma_{e} - 1)$,
$\gamma_{e} ^{\prime} \sim 1.1 > \gamma_{e, \mathrm{Comp}}$.
Thus, Comptonization is still important for the
$\tau_{\mathrm{initial}} = 16$ simulations and the photons are still
gaining energy from the electrons. The photons will gain energy from
the electrons until they reach the energy of the electrons. For the
seed $f_{\nu} \propto \nu^{-1/2}$ spectrum, initially there are more
photons at lower energies near $E_{1, \gamma}$. However, as
$\tau_{\mathrm{initial}}$ increases, the additional scatterings allow
for more photons at lower energies to be upscattered to higher
energies and they also allow for more photons to reach the energy of
the electrons. The spectrum flattens because more photons are being
removed from lower energies and being placed at high energies near the
energy of the electrons. If we considered $\tau_{\mathrm{initial}} > 16$,
if Compton-$Y$ remains greater than 1, eventually all the photons will
be upscattered to energies close to the energy of the electrons. The
photon spectrum will no longer look flat, but will instead be peaked
at the energy of the electrons.

Lastly, we note that although the simulation output spectrum
for $\tau_{\mathrm{initial}} = 16$ in Figure \ref{fig8}
is very similar to the low-energy spectrum for the
prompt emission, these simulation results cannot explain it. The spectrum peaks at
the energy of the electrons, which have an energy
$\sim 50 \mbox{ keV}$ in the jet-comoving frame and thus an energy
$\sim 50 \mbox{ keV} \times 300 \sim 10^{4} \mbox{ keV}$
in the observer frame ($\Gamma=300$). In order for the spectrum to
break at $\sim 300 \mbox{ keV}$ in the observer frame,
the electrons would need to have an
energy $\sim 1 \mbox{ keV}$ in the jet-comoving frame.
In addition, once the $f_{\nu} \propto \nu^{0}$ spectrum ends in 
Figure \ref{fig8}, the spectrum declines very rapidly, in disagreement
with the observed $f_{\nu} \propto \nu^{-1.2}$ for the prompt emission. 
%
%
%

\section{Conclusions}
\label{conclusions_sect}

In this work, we presented our methodology for our MC photospheric
code and our simulation results for a wide parameter space
with a realistic photon to electron
ratio $N_{\gamma} / N_{e} = 10^{5}$, as expected for the GRB prompt
emission. Our goal was to determine if the photospheric process can
explain the observed high-energy spectrum
$f_{\nu} \propto \nu^{-1.2}$ of the prompt emission. For these simulations, we considered
the Comptonization of a seed BB spectrum. If electron re-heating is
not considered, we determined that
considering both low photon temperatures and large
optical depths $\sim 10-20$ is best for producing a power-law spectrum above the BB
peak energy for the following two reasons: 1. low temperature photons
cool electrons more slowly, allowing more photons to be upscattered to
higher energies. 2. At larger optical depths, the average number of
scatterings a photon experiences is larger, allowing for more photons
to be upscattered to higher energies. On the other hand, the output
spectrum for the cases we considered with higher photon temperatures
and low optical depths display a sharp drop in $f_{\nu}$ above the
peak energy by $\sim 2$ orders of magnitude.
These cases require additional energy in the
electrons and we
demonstrated that if we consider electron re-heating, a power-law
spectrum above the peak-energy can be produced.

One issue with the simulations with a low photon temperature
and a large optical depth is that it is difficult to match the
peak-energy of the prompt emission since adiabatic cooling of
photons decreases the BB peak-energy by a large factor for large
optical depths. For these simulations 
(lower-left panel of \ref{fig4}), the spectrum peaks at 
$\sim \mbox{30 keV}$, 10 times lower than the typical peak-energy of
the prompt emission. On the other hand, when considering electron
re-heating, we find simulation results with 1. peak-energy $\sim
\mbox{100 keV}$ 2. a power-law spectrum extending to at least 10 MeV
3. a spectral index similar to $f_{\nu} \propto \nu^{-1.2}$, all in
agreement with the prompt emission observations.
Thus, considering electron re-heating near the
photosphere with a large photon temperature
($\sim \mbox{ 100 keV}$ in the observer frame) may be the best
solution for explaining the prompt emissions observations
with the photospheric process. We note that the electron
  re-heating scheme we consider is a global re-heating mechanism since
we re-accelerate all the electrons in our simulations after a
re-heating episode. In GRBs, the heating mechanism is likely to be
global, i.e. occurring throughout the causally connected part of the
jet, since observations show that the conversion of jet energy to
gamma-ray radiation is an efficient process.

We also performed photospheric simulations for different
values of $N_{\gamma}/N_{e}$ and demonstrated that the simulation
results have a strong dependence on the photon to electron
ratio. This result highlights the important of performing realistic
photospheric simulations with $N_{\gamma}/N_{e} = 10^{5}$.

In addition, we also used our MC photospheric code to study
how Comptonization modifies a $f_{\nu} \propto \nu^{-1/2}$ seed spectrum, as
expected for synchrotron when electrons are in the fast cooling
regime. For large optical depths, the effect is to flatten the spectrum
to $f_{\nu} \propto \nu^{0}$, reminiscent of the low-energy spectrum
of the Band function. However, these simulation results cannot
explain the low-energy spectrum of the prompt emission since the
simulation output spectrum peaks at $\sim 10^{4} \mbox{ keV}$, much larger
than the $\sim 300 \mbox{ keV}$ peak energy of the prompt emission.
In addition, above the peak-energy, the spectrum declines very
rapidly, in disagreement with the observed $f_{\nu} \propto \nu^{-1.2}$
for the prompt emission.
%
%
%

\section*{Acknowledgements}
\noindent
We thank the referee for a prompt and constructive referee report,
which significantly improved the manuscript.
The authors would like to thank Volker Bromm and
Milo\v{s} Milosavljevi\'{c} for generously providing computational resources.
R.S. would like to dedicate this work to Rodolfo Barniol Duran, who
has been a good mentor and who has taught me a lot about GRBs.
R.S. also thanks Davide Lazzati and Wenbin Lu for helpful discussions.




\bibliographystyle{mnras}




\appendix

\section{Conventions for Appendices}

In the appendices below, we describe the details of the algorithm we
adopted for our MC Photospheric code. The algorithms in
Appendices \ref{appendix_elec_dir}, \ref{appendix_elec_gam_e_MB_PL},
\ref{appendix_phot_dir}, \ref{appendix_phot_ener}, and \ref{appen_elec_phot_scatt}
were adopted from the comprehensive IC$/$Compton scattering
reference \citet{pozdnyakov_et_al_1983}.

The conventions we describe here apply to all the Appendices below.
All primed (un-primmed) quantities are in the jet-comoving (observer)
frame. A quantity with a subscript $i$ (initial) [$f$ (final)]
refers to the quantity before [after] the scattering event.
All the random numbers we draw are denoted by $\xi$
and they are uniform random numbers in the
interval 0 to 1. When referring to a vector $\mathbf{A}$, the vector is
put in boldface and the magnitude of the vector is denoted by
$\| \mathbf{A} \|$. Lastly, $\hat{x}$, $\hat{y}$, $\hat{z}$ represent the unit vectors
in Cartesian coordinates.
\section{Initialization of Electrons}
\subsection{Drawing Random Electron Directions}
\label{appendix_elec_dir}

The momentum of an electron is given by
$\mathbf{p} _{e} ^{\prime} = \| \mathbf{p} _{e} ^{\prime} \|
\mathrm{v}_{1} ^{\prime} \hat{x} +
\| \mathbf{p} _{e} ^{\prime} \| \mathrm{v}_{2} ^{\prime} \hat{y}+ \|
\mathbf{p} _{e} ^{\prime} \| \mathrm{v}_{3} ^{\prime} \hat{z}$ ,
where $\mathrm{v}_{1} ^{\prime}$, $\mathrm{v}_{2} ^{\prime}$, and
$\mathrm{v}_{3} ^{\prime}$ are the
components of $\mathbf{p} _{e} ^{\prime}$ in the x, y, and z
directions, respectively. To draw a random direction for an
electron, draw two random numbers
$\xi_{1}$, $\xi_{2}$ and use the expressions
\begin{eqnarray}
\mathrm{v}_{3} ^{\prime} &=& 2 \xi_{1} - 1 \nonumber \\
\mathrm{v}_{2} ^{\prime} &=& \sqrt{1 - (\mathrm{v}_{3} ^{\prime})^{2} } \sin{( 2 \pi \xi_{2} )} \nonumber \\
\mathrm{v}_{1} ^{\prime} &=& \sqrt{1 - (\mathrm{v}_{3} ^{\prime})^{2} } \cos{( 2 \pi \xi_{2} ) } . \nonumber
\end{eqnarray}
\subsection{Drawing Electron Energy from MB and PL Distributions}
\label{appendix_elec_gam_e_MB_PL}
\noindent
\subsubsection{Maxwell Boltzmann Electrons}

The dimensionless energy ($n$) and the dimensionless momentum ($\eta$)
of an electron are given by $n = k_{B} T_{e} ^{\prime} / ( m_{e} c^{2}) $
and $\eta = \| \mathbf{p} _{e} ^{\prime} \|/(m_{e} c) $,
where $\| \mathbf{p} _{e} ^{\prime} \| = m_{e} c \beta_{e} ^{\prime} \gamma_{e} ^{\prime}$.
From the expression for $\eta$,
$\gamma_{e} ^{\prime} = \sqrt{ \eta^{2} + 1 }$ and
$\beta_{e} ^{\prime} = \eta/\gamma_{e} ^{\prime}$. \citet{pozdnyakov_et_al_1983} present two
algorithms for drawing $\gamma_{e} ^{\prime}$ for an electron,
depending on whether $ k_{B} T_{e} ^{\prime}$ is less than or larger
than 150 keV.
If $ k_{B} T_{e} ^{\prime} < 150 \mbox{ keV}$, draw two
random numbers $\xi_{1}$, $\xi_{2}$. First,  calculate
$\xi^{\prime}$ with the expression
$\xi^{\prime} = -(3/2)\mbox{ ln}(\xi_{1})$. Then, test the
acceptance-rejection condition
$\xi_{2} ^{2} < 0.151( 1 + n \xi^{\prime} )^{2} \xi^{\prime} (2 + n \xi^{\prime}) \xi_{1} $.
If it is satisfied, set
$\eta = \sqrt{ n \xi^{\prime} ( 2 + n \xi^{\prime} )} $,
$\gamma_{e} ^{\prime} = \sqrt{ \eta^{2} + 1 }$, and
$\beta_{e} ^{\prime} = \eta/\gamma_{e} ^{\prime}$. Otherwise, continue
to draw new $\xi_{1}$, $\xi_{2}$ until the acceptance-rejection
condition is satisfied.

If $ k_{B} T_{e} ^{\prime} \ge 150 \mbox{ keV}$, draw four
random numbers $\xi_{1}$, $\xi_{2}$, $\xi_{3}$, $\xi_{4}$.
Compute the quantities
$ \eta ^{\prime} = - n \mbox{ ln} (\xi_{1} \xi_{2} \xi_{3}) $ and
$ \eta ^{\prime \prime} = - n \mbox{ ln} (\xi_{1} \xi_{2} \xi_{3} \xi_{4}) $
and test the acceptance-rejection condition
$ (\eta ^{\prime \prime})^{2} - (\eta ^{\prime})^{2} > 1 $.
If it is satisfied, set
$ \eta =  \eta ^{\prime} $,
$\gamma_{e} ^{\prime} = \sqrt{ \eta^{2} + 1 }$, and
$\beta_{e} ^{\prime} = \eta/\gamma_{e} ^{\prime}$. Otherwise, continue
to draw new $\xi_{1}$, $\xi_{2}$, $\xi_{3}$, $\xi_{4}$
until the acceptance-rejection condition is satisfied.
\subsubsection{Power-Law Distribution of Electrons}

To drawn $\gamma_{e} ^{\prime}$ for an electron following a PL
distribution, draw one random number $\xi_{1}$
and calculate $E^{\prime} = m_{e} c^{2} \gamma_{e} ^{\prime}$ with the expression
\begin{equation}
E^{\prime} = [\xi_{1} ((E_{2} ^{\prime}) ^{1-p} - (E_{1} ^{\prime})
^{1-p}) + (E_{1} ^{\prime}) ^{1-p}  ]^{1/(1-p)} . \nonumber
\end{equation}
Then, we set $\gamma_{e} ^{\prime} = E^{\prime}/(m_{e} c^{2})$ and
$\beta_{e} = \sqrt{1 - (\gamma_{e} ^{\prime})^{-2} }$.
\section{Initialization of Photons}
\subsection{Photon Directions}
\label{appendix_phot_dir}

The momentum of a photon is given by
$\mathbf{p} _{\gamma} ^{\prime} = (E_{\gamma} ^{\prime}/c) \Omega_{1} ^{\prime} \hat{x} +
 (E_{\gamma} ^{\prime}/c) \Omega_{2} ^{\prime} \hat{y} +
 (E_{\gamma} ^{\prime}/c) \Omega_{3} ^{\prime} \hat{z} $,
where $\Omega_{1} ^{\prime}$, $\Omega_{2} ^{\prime}$, and
$\Omega_{3} ^{\prime}$ are the components of $\mathbf{p} _{\gamma} ^{\prime}$
in the x, y, and z directions, respectively. To draw a random direction for a
photon, draw two random numbers $\xi_{1}$, $\xi_{2}$ and use
the expressions (same algorithm as in Appendix \ref{appendix_elec_dir})
\begin{eqnarray}
\Omega_{3} ^{\prime} &=& 2 \xi_{1} - 1 \nonumber \\
\Omega_{2} ^{\prime} &=& \sqrt{1 - (\Omega_{3} ^{\prime})^{2} } \sin{ ( 2 \pi \xi_{2} ) } \nonumber \\
\Omega_{1} ^{\prime} &=& \sqrt{1 - (\Omega_{3} ^{\prime})^{2} } \cos{ (2 \pi \xi_{2} ) } . \nonumber
\end{eqnarray}
\subsection{Photon Energies}
\label{appendix_phot_ener}
\subsubsection{BB Distribution}
\noindent
Draw four random numbers $\xi_{1}$, $\xi_{2}$, $\xi_{3}$, $\xi_{4}$.
Then, determine $\alpha$ such that
\begin{equation*}
\alpha = \left\{
\begin{array}{ll}
\hskip -7pt 1 \mbox{ if }
 & 1.202 \xi_{1} < 1 \nonumber \\
\hskip -7pt m \mbox{ if }
& \sum\limits_{1}^{m-1} j^{-3} \le 1.202 \xi_{1} < \sum\limits_{1}^{m}
  j^{-3} .
\end{array}
\right . \nonumber
\end{equation*}
Then, set $E_{\gamma} ^{\prime} = - (k_{B} T_{\gamma} ^{\prime}/\alpha)
\mbox{ ln} (\xi_{2} \xi_{3} \xi_{4} )$.
\subsubsection{Power-Law Distribution}
\noindent
To draw a photon power-law distribution with
$E_{\gamma, 1} ^{\prime} < E < E_{\gamma, 2} ^{\prime}$,
draw one random number $\xi_{1}$ and set
\begin{equation}
E_{\gamma} ^{\prime} = [\xi_{1} ((E_{\gamma, 2} ^{\prime}) ^{1-p} - (E_{\gamma, 1} ^{\prime})
^{1-p}) + (E_{\gamma, 1} ^{\prime}) ^{1-p}  ]^{1/(1-p)} . \nonumber
\end{equation}
The $f_{\nu}$ photon spectrum is given by $f_{\nu} \propto \nu^{1-p}$.
\subsection{Photon Propagation}
\label{appendix_phot_propagation}

In this subappendix, we calculate the new position of
a photon in the observer frame after it has traveled a distance
$s^{\prime}$ in the jet-comoving frame. We denote the initial [final] position
of the photon in the observer frame as
$\mathbf{R_{i}} = R_{1,i} \hat{x} +  R_{2,i} \hat{y} +  R_{3,i} \hat{z}$
[$\mathbf{R_{f}} = R_{1,f} \hat{x} +  R_{2,f} \hat{y} +  R_{3,f} \hat{z}$].
The time traveled by the
photon in the jet-comoving frame ($\Delta t ^{\prime}$) and the
displacements of the photon in the jet-comoving frame in the
x ($\Delta R_{1} ^{\prime}$), y ($\Delta R_{2} ^{\prime}$),
and z ($\Delta R_{3} ^{\prime}$) directions are given by
\begin{eqnarray}
\Delta t ^{\prime}  &=& s^{\prime} (R) /c   \nonumber \\
\Delta R_{1} ^{\prime}  &=& s^{\prime} (R) \Omega_{1,i} ^{\prime}  \nonumber \\
\Delta R_{2} ^{\prime}   &=& s^{\prime} (R) \Omega_{2,i} ^{\prime} \nonumber \\
\Delta R_{3} ^{\prime}  &=& s^{\prime} (R) \Omega_{3,i} ^{\prime} , \nonumber
\end{eqnarray}
where $\Omega_{1,i} ^{\prime}$ $ ( \Omega_{2,i} ^{\prime} )$
$[ \Omega_{3,i} ^{\prime}]$ is the direction of the photon
in the x (y) $[$z$]$ direction before the scattering event.

\noindent
From the bulk Lorentz factor of the jet, the speed of the jet divided
by the speed of light is $\beta_{j} = \sqrt{1 - \Gamma^{-2}}$. The
components of $\beta_{j}$ in the x ($\beta_{j, 1}$),
y ($\beta_{j, 2}$), and z ($\beta_{j, 3}$) directions are given by
\begin{eqnarray}
\beta_{j, 1}  &=& \beta_{j} \cos{\phi} \sin{\theta}  \nonumber \\
\beta_{j, 2}  &=& \beta_{j} \sin{\phi} \sin{\theta}  \nonumber \\
\beta_{j, 3}  &=& \beta_{j} \cos{\theta} .  \nonumber
\end{eqnarray}
The angles $\theta$ and $\phi$, corresponding to the photon position in
spherical coordinates, can be found with the expressions
$\cos{\theta} = R_{3,i}/\sqrt{(R_{1,i})^{2} + (R_{2,i})^{2} + (R_{3,i})^{2}}$
and $\tan{\phi} =  R_{2,i} / R_{1,i}$. With these results, we can
Lorentz transform the displacements from the jet comoving frame to the
observe frame with the expressions
\begin{eqnarray}
\Delta R_{1}  &=& \Gamma \beta_{j, 1} c \Delta t ^{\prime}
+ \left[ 1 + (\Gamma-1) \frac{\beta_{j, 1} ^{2}}{\beta_{j} ^{2}} \right] \Delta R_{1} ^{\prime} +
\nonumber \\
&& \left[  (\Gamma -1) \frac{\beta_{j, 1} \beta_{j, 2} }{\beta_{j} ^{2}} \right] \Delta R_{2} ^{\prime} +
\left[  (\Gamma -1) \frac{\beta_{j, 1} \beta_{j, 3} }{\beta_{j} ^{2}} \right] \Delta R_{3} ^{\prime}
\nonumber \\
\Delta R_{2}  &=& \Gamma \beta_{j, 2} c \Delta t ^{\prime}
+ \left[  (\Gamma -1) \frac{\beta_{j, 2} \beta_{j, 1} }{\beta_{j} ^{2}} \right] \Delta R_{1} ^{\prime} +
\nonumber \\
&& \left[ 1 + (\Gamma-1) \frac{\beta_{j, 2} ^{2}}{\beta_{j} ^{2}} \right] \Delta R_{2} ^{\prime}
+  \left[  (\Gamma -1) \frac{\beta_{j, 2} \beta_{j, 3} }{\beta_{j} ^{2}} \right] \Delta R_{3} ^{\prime}
\nonumber \\
\Delta R_{3}  &=& \Gamma \beta_{j, 3} c \Delta t ^{\prime}
+ \left[  (\Gamma -1) \frac{\beta_{j, 3} \beta_{j, 1} }{\beta_{j} ^{2}} \right] \Delta R_{1} ^{\prime} +
\nonumber \\
&& \left[  (\Gamma -1) \frac{\beta_{j, 3} \beta_{j, 2} }{\beta_{j} ^{2}} \right] \Delta R_{2} ^{\prime}
+ \left[ 1 + (\Gamma-1) \frac{\beta_{j, 3} ^{2}}{\beta_{j} ^{2}} \right] \Delta R_{3} ^{\prime} .
\nonumber \\
\Delta t  &=& \Gamma \Delta t ^{\prime} +
\Gamma \beta_{j, 1} ( \Delta R_{1} ^{\prime} / c ) +
\nonumber \\
&& \Gamma \beta_{j, 2} ( \Delta R_{2} ^{\prime} / c ) +
\Gamma \beta_{j, 3} ( \Delta R_{3} ^{\prime} / c ) .
\nonumber
\end{eqnarray}
In the above equation, $\Delta t$ is the time it took the photon to
travel a distance $s^{\prime}$ in the observer frame.
The new position of the photon and the new
distance of the photon from the central engine
can be found with the equations
\begin{eqnarray}
R_{1,f}  &=& R_{1,i} + \Delta R_{1} \nonumber \\
R_{2,f}  &=& R_{2,i} + \Delta R_{2} \nonumber \\
R_{3,f}  &=& R_{3,i} + \Delta R_{3}  \nonumber \\
R  &=& \sqrt{ (R_{1,f})^{2} + (R_{2,f})^{2} + (R_{3,f})^{2} }  . \nonumber
\end{eqnarray}
\section{Electron Photon Scattering Interaction}
\label{appen_elec_phot_scatt}

In this Appendix, we describe the algorithm from Chapter 9 of
\citet{pozdnyakov_et_al_1983} for calculating the final
energy $(E_{\gamma, f} ^{\prime})$ and direction of a photon
$(\Omega_{1,f} ^{\prime} , \Omega_{2,f} ^{\prime} ,
\Omega_{3,f} ^{\prime})$ after a scattering event.
The angle between the electron direction and photon direction before the scattering
event is given by
\begin{equation*}
\mu_{i} ^{\prime} = \mathrm{v}_{1,i} ^{\prime} \Omega_{1,i} ^{\prime}  +
\mathrm{v}_{2,i} ^{\prime} \Omega_{2,i} ^{\prime} +
\mathrm{v}_{3,i} ^{\prime} \Omega_{3,i} ^{\prime} . \nonumber
\end{equation*}
With the angle $\mu_{i} ^{\prime} $, we can compute the dimensionless
photon energy in the electron rest-frame, denoted as $x_{i} ^{\prime}$,
 with the expression
\begin{equation*}
x_{i} ^{\prime} = 2 \gamma_{e , i} ^{\prime} (E_{\gamma, i} ^{\prime} / m_{e} c^{2}
) ( 1 - \mu_{i} ^{\prime} \beta_{e,i} ^{\prime} ) . \nonumber
\end{equation*}
\noindent
With $x_{i} ^{\prime}$, we can compute the cross-section for this interaction,
$\sigma (x_{i} ^{\prime}) = 2 \pi r_{e} ^{2} \hat{\sigma} (x_{i} ^{\prime})$ ,
where $r_{e} = e^{2}/(m_{e} c^{2})$ is the Classical electron radius,
$e$ is the electron charge, and $\hat{\sigma}$ is calculated
with the expression \citep{pozdnyakov_et_al_1983}
\begin{equation*}
\hat{\sigma} (x_{i} ^{\prime}) = \left\{
\begin{array}{ll}
\hskip -7pt  1/3 + 0.141 x_{i} ^{\prime} - 0.12 (x_{i} ^{\prime})^{2}
+ & \nonumber \\
( 1 + 0.5 x_{i} ^{\prime} ) ( 1 + x_{i} ^{\prime} )^{-2},  &
x_{i} ^{\prime} \le 0.5  ; \nonumber \\
\hskip -7pt  [ \mbox{ln} (1 + x_{i} ^{\prime}) + 0.06] (x_{i} ^{\prime})^{-1} , &
0.5 \le  x_{i} ^{\prime} \le 3.5  ; \nonumber \\
\hskip -7pt  [ \mbox{ln} (1 + x_{i} ^{\prime}) + 0.5 -
(2 + 0.076 x_{i} ^{\prime}) ^{-1} ] (x_{i} ^{\prime})^{-1} , &
3.5 \le  x_{i} ^{\prime}  . \nonumber
\end{array}
\right . \nonumber
\end{equation*}
To test if the scattering event
will occur, we draw a random number $\xi_{s}$.
If $\xi_{s} < \sigma(x_{i} ^{\prime})/\sigma_{T}$ is satisfied, the scattering
event occurs.

If the scattering event is determined to occur,
we perform an acceptance-rejection for the direction
and energy of the photon after the scattering event.
We first draw 3 random numbers $\xi_{1}$, $\xi_{2}$, $\xi_{3}$.
The random numbers $\xi_{1}$, $\xi_{2}$ are used to compute the angles
\begin{eqnarray}
\mu_{f} ^{\prime} &=& \frac{\beta_{e,i} ^{\prime} + 2 \xi_{1} - 1 }{1 +
            \beta_{e,i} ^{\prime} ( 2 \xi_{1} - 1)} \nonumber \\
\phi_{f} ^{\prime} &=& 2 \pi \xi_{2} . \nonumber
\end{eqnarray}
From $\mu_{f} ^{\prime}$, $\phi_{f} ^{\prime}$, we can compute the
direction of the photon after the scattering event with the expressions
\begin{eqnarray}
\Omega_{1,f} ^{\prime} &=& \mu_{f} ^{\prime} \mathrm{v}_{3,i} ^{\prime} +
\sqrt{1 - (\mu_{f} ^{\prime}) ^{2} } \rho^{-1} ( \mathrm{v}_{2,i} ^{\prime} \cos{\phi_{f}} +
\mathrm{v}_{1,i} ^{\prime} \mathrm{v}_{3,i} ^{\prime} \sin{\phi_{f} ^{\prime}} ) \nonumber \\
\Omega_{2,f} ^{\prime} &=& \mu_{f} ^{\prime} \mathrm{v}_{2,i} ^{\prime} +
\sqrt{1 - (\mu_{f} ^{\prime}) ^{2} } \rho^{-1} ( - \mathrm{v}_{1,i}
^{\prime} \cos{\phi_{f} ^{\prime}} +
\mathrm{v}_{2,i} ^{\prime} \mathrm{v}_{3,i} ^{\prime} \sin{\phi_{f} ^{\prime}} ) \nonumber \\
\Omega_{3,f} ^{\prime} &=& \mu_{f} ^{\prime} \mathrm{v}_{3,i} ^{\prime} +
\sqrt{1 -(\mu_{f} ^{\prime}) ^{2} } \rho \sin{\phi_{f} ^{\prime}} , \nonumber
\end{eqnarray}
where $ \rho = \sqrt{ (\mathrm{v}_{1,i} ^{\prime})^{2} + (\mathrm{v}_{2,i} ^{\prime})^{2}} $.
The next step is to compute the ratio of the dimensionless photon energy in
the electron rest frame after scattering $(x_{f} ^{\prime})$, to $x_{i}
^{\prime}$, with the equation
\begin{eqnarray}
\frac{x_{f} ^{\prime}}{x_{i} ^{\prime}} &=& \left[ 1 +
\frac{E_{\gamma, i} ^{\prime}( 1 -\mathbf{\Omega_{i} ^{\prime}} \cdot
\mathbf{\Omega_{f} ^{\prime}} ) }{\gamma_{e,i} ^{\prime} m_{e} c^{2}
(1 - \mu_{f} \beta_{e, i} ^{\prime} ) } \right]^{-1} , \nonumber
\end{eqnarray}
where
$
\mathbf{\Omega_{i} ^{\prime}} \cdot \mathbf{\Omega_{f} ^{\prime}}
        = \Omega_{1,i} ^{\prime} \Omega_{1,f} ^{\prime} +
                \Omega_{2,i} ^{\prime} \Omega_{2,f} ^{\prime} +
                 \Omega_{3,i} ^{\prime} \Omega_{3,f} ^{\prime}
$
is the angle between the initial photon direction and the
final photon direction.
To determine if the final photon energy and direction is accepted, we
use $\xi_{3}$ to test the acceptance-rejection condition
\begin{eqnarray}
2 \xi_{3} &<& \left( \frac{x_{f} ^{\prime}}{x_{i} ^{\prime}} \right)^{2}
            \mathcal{X} \mbox{ , where } \nonumber \\
\mathcal{X} &=& \left( \frac{x_{f} ^{\prime}}{x_{i} ^{\prime}} \right)^{-1}
         + \frac{x_{f} ^{\prime}}{x_{i} ^{\prime}}
         + \frac{4}{x_{i} ^{\prime}} \left[ 1 - \left( \frac{x_{f} ^{\prime}}{x_{i} ^{\prime}} \right)^{-1}  \right]
         + \frac{4}{(x_{i} ^{\prime})^{2}} \left[ 1 - \left( \frac{x_{f} ^{\prime}}{x_{i} ^{\prime}} \right)^{-1}  \right]^{2} .
\nonumber
\end{eqnarray}
If the condition $2 \xi_{3} <(x_{f} ^{\prime}/x_{i} ^{\prime})^{2} \mathcal{X} $
is satisfied, the final photon direction is accepted and the energy of
the photon after scattering is found with the expression
$
E_{\gamma, f} ^{\prime} =
(x_{f} ^{\prime}/x_{i} ^{\prime}) x_{i} ^{\prime} m_{e} c^{2}
[2 \gamma_{e,i} ^{\prime} (1 - \mu_{f} ^{\prime} \beta_{e,i} ^{\prime} ) ]^{-1} $.
Otherwise, we continue to draw 3 new random numbers
$\xi_{1}$, $\xi_{2}$, $\xi_{3}$ until the condition
$2 \xi_{1} <(x_{f} ^{\prime}/x_{i} ^{\prime})^{2} \mathcal{X} $
is satisfied.
\section{Updating Electron Energy and Direction After Scattering}
\label{appen_elec_ener_dir_update}

In addition to updating the energy and direction of the photon after
the scattering event, we also update the energy and direction of the
electron. The energy of a photon-electron
system is given by
$E^{\prime} = E _{\gamma} ^{\prime} + m_{e} c^{2} \gamma_{e} ^{\prime}$.
Using conservation of energy, we can find $\gamma_{e, f} ^{\prime}$
with the expression
\begin{eqnarray}
\gamma_{e,f} ^{\prime} &=&\frac{E_{\gamma, i} ^{\prime} - E_{\gamma, f} ^{\prime}
+ m_{e} c^{2} \gamma_{e,i} ^{\prime} }{m_{e} c^{2} } . \nonumber
\end{eqnarray}
The momentum of a photon-electron system is given by
$\mathbf{p} ^{\prime} = [( E _{\gamma} ^{\prime}/c) \Omega_{1} ^{\prime} +
\| \mathbf{p} _{e} ^{\prime} \| \mathrm{v}_{1} ^{\prime} ] \hat{x} +
 [( E _{\gamma} ^{\prime}/c) \Omega_{2} ^{\prime} +
\| \mathbf{p} _{e} ^{\prime} \| \mathrm{v}_{2} ^{\prime} ] \hat{y} +
[( E _{\gamma} ^{\prime}/c) \Omega_{3} ^{\prime} +
\| \mathbf{p} _{e} ^{\prime} \| \mathrm{v}_{3} ^{\prime} ] \hat{z} $, where
$\| \mathbf{p} _{e} ^{\prime} \| = m_{e} c \beta_{e} ^{\prime} \gamma_{e} ^{\prime}$.
Using conservation of momentum in each direction,
we can find $\mathrm{v}_{1,f} ^{\prime}$, $\mathrm{v}_{2,f} ^{\prime}$,
$\mathrm{v}_{3,f} ^{\prime}$ with the expressions
\begin{eqnarray}
\mathrm{v}_{1,f} ^{\prime} &=& \frac{(E_{\gamma, i} ^{\prime}/c) \Omega_{1,i} ^{\prime}  +
m_{e} c \beta_{e, i} ^{\prime} \gamma_{e, i} ^{\prime} \mathrm{v}_{1,i} ^{\prime} -
(E_{\gamma, f} ^{\prime}/c) \Omega_{1,f} ^{\prime} }{m_{e} c \beta_{e, f} ^{\prime} \gamma_{e, f} ^{\prime}} \nonumber \\
\mathrm{v}_{2,f} ^{\prime} &=& \frac{(E_{\gamma, i} ^{\prime}/c) \Omega_{2,i} ^{\prime}  +
m_{e} c \beta_{e, i} ^{\prime} \gamma_{e, i} ^{\prime} \mathrm{v}_{2,i} ^{\prime} -
(E_{\gamma, f} ^{\prime}/c) \Omega_{2,f} ^{\prime} }{m_{e} c \beta_{e, f} ^{\prime} \gamma_{e, f} ^{\prime}} \nonumber \\
\mathrm{v}_{3,f} ^{\prime} &=& \frac{(E_{\gamma, i} ^{\prime}/c) \Omega_{3,i} ^{\prime}  +
m_{e} c \beta_{e, i} ^{\prime} \gamma_{e, i} ^{\prime} \mathrm{v}_{3,i} ^{\prime} -
(E_{\gamma, f} ^{\prime}/c) \Omega_{3,f} ^{\prime} }{m_{e} c \beta_{e, f} ^{\prime} \gamma_{e, f} ^{\prime}} . \nonumber
\end{eqnarray}


\bsp	
\label{lastpage}
\end{document}